\documentclass[fleqn,usenatbib]{mnras}

\usepackage{newtxtext,newtxmath}

\usepackage[T1]{fontenc}

\DeclareRobustCommand{\VAN}[3]{#2}
\let\VANthebibliography\thebibliography
\def\thebibliography{\DeclareRobustCommand{\VAN}[3]{##3}\VANthebibliography}

\usepackage{graphicx}	
\usepackage{amsmath}	
\usepackage{amssymb}	
\usepackage{siunitx}
\usepackage{subfigure}

\newcommand{\mystar}{WASP-94A}
\renewcommand{\th}{\textsuperscript{th}}
\newcommand{\lrhk}{\ensuremath{\log R'_{\rm HK}}}

\title[Transmission spectrum of WASP-94A\,b]{
LRG-BEASTS: sodium absorption and Rayleigh scattering in the atmosphere of WASP-94A\,b using NTT/EFOSC2}

\author[E. Ahrer et al.]{
E. Ahrer$^{1,2}$\thanks{E-mail: eva-maria.ahrer@warwick.ac.uk},
P. J. Wheatley$^{1,2}$\thanks{E-mail: p.j.wheatley@warwick.ac.uk},
J. Kirk$^{3}$, S. Gandhi$^{4,1,2}$, G. W. King$^{5,1,2}$, 
T. Louden$^{1,2}$
\\
$^{1}$Centre for Exoplanets and Habitability, University of Warwick, Gibbet Hill Road, CV4 7AL Coventry, UK\\
$^{2}$Department of Physics, University of Warwick, Gibbet Hill Road, CV4 7AL Coventry, UK\\
$^{3}$Center for Astrophysics | Harvard \& Smithsonian, 60 Garden Street, Cambridge, MA 02138, USA\\
$^{4}$Leiden Observatory, Leiden University, Postbus 9513, 2300 RA Leiden, The Netherlands\\
$^{5}$Department of Astronomy, University of Michigan, Ann Arbor, MI 48109, USA\\
}

\date{Accepted XXX. Received YYY; in original form ZZZ}

\pubyear{2021}

\begin{document}
\label{firstpage}
\pagerange{\pageref{firstpage}--\pageref{lastpage}}
\maketitle

\begin{abstract}
We present an optical transmission spectrum for \mystar\,b, the first atmospheric characterisation of this highly-inflated hot Jupiter.  The planet has a reported radius of $1.72^{+0.06}_{-0.05}$~R$_{\textrm{Jup}}$, a mass of only $0.456^{+0.032}_{-0.036}$~M$_{\textrm{Jup}}$, and an equilibrium temperature of $1508 \pm 75$~K. We observed the planet transit spectroscopically with the EFOSC2 instrument on the ESO New Technology Telescope (NTT) at La Silla, Chile: 
the first use of NTT/EFOSC2 for transmission spectroscopy.
We achieved an average transit-depth precision of $128$~ppm for bin widths of $\sim200$\,\AA.
This high precision was achieved in part by linking Gaussian Process hyperparameters across all wavelength bins. 
The resulting transmission spectrum, spanning a wavelength range of $3800 - 7140$\,\AA, exhibits a sodium absorption with a significance of $4.9\sigma$, suggesting a relatively cloud-free atmosphere. The sodium signal may be broadened, with a best fitting width of $78_{-32}^{+67}$\,\AA\ in contrast to the instrumental resolution of $27.2 \pm 0.2$\,\AA. We also detect a steep slope in the blue end of the transmission spectrum, indicating the presence of Rayleigh scattering in the atmosphere of \mystar\,b. Retrieval models show evidence for the observed slope to be super-Rayleigh and potential causes are discussed. Finally, we find narrow absorption cores in the CaII H\&K lines of \mystar, suggesting the star is enshrouded in gas escaping the hot Jupiter. 
\end{abstract}




\section{Introduction}


The field of atmospheric characterisation of exoplanets has developed rapidly 
since the first detection of an exoplanet atmosphere by \citet{Charbonneau2002DetectionAtmosphere}, 
where
sodium absorption was found
in the atmosphere of HD~209458\,b
using the Hubble Space Telescope 
\citep[HST; although this detection has recently been called into question][]{Casasayas-Barris2020IsStudies,Casasayas-Barris2021TheESPRESSO}.
The existence of sodium in the atmosphere of hot exoplanets had been previously predicted by \citet{Seager2000TheoreticalTransits} as atomic sodium can exist in the gas phase at high temperatures. Since then, other hot close-in exoplanets have been found to have sodium absorption in their atmospheric spectra \citep[e.g.\ see summary in][]{Madhusudhan2019ExoplanetaryProspects}. However, there are still fewer than thirty  sodium detections overall, and the majority have been made with high-resolution spectrographs, the first examples being \citet{Redfield2008SodiumSpectrum} and \citet{Snellen2008Ground-based209458b}.  Low-resolution ground-based detections remain challenging and only a handful have been made to date \citep[e.g.][]{Sing2012GTCSpectroscopy,Nikolov2018AnExoplanet,Nikolov2016VLTGROUND,Alderson2020LRG-BEASTS:WASP-21b}.

Soon after the first detection of sodium, carbon and oxygen were found 
escaping the atmosphere of HD\,209458\,b using ultraviolet spectra \citep{Vidal-Madjar2004Detection209458b},
while other atomic species such as potassium were found later in the atmosphere of hot Jupiters at optical wavelengths \citep[e.g.][]{Sing2011GranSpectrophotometry, Nikolov2015HSTWASP-6b, Sing2015HSTScattering,Sing2016ADepletion,Chen2018TheWASP-127b}. In addition, infrared observations with both space and ground-based telescopes e.g.\ the HST and the Very Large Telescope (VLT) commenced detections of molecular bands such as water and carbon monoxide \citep[e.g.][]{Snellen2010The209458b,Brogi2012TheB, Birkby2013Detection3.2m,Deming2013InfraredTelescope, Wakeford2013HSTObservations,Kreidberg2014AWASP-43b,Hawker2018Evidence209458b}.

Non-detections of molecular and/or atomic species can be of equal importance as they provide information on the presence of clouds and hazes in the planetary atmosphere \citep[e.g.][]{Gibson2013ANM,Knutson2014AGJ436b,Kreidberg2014Clouds1214b,Louden2017AWASP-52b, Spyratos2021TransmissionWASP-88b}. Studies across samples of planets can attempt to identify the conditions for cloud and haze production, which are potentially linked to the equilibrium temperature and/or surface gravity of the exoplanet \citep[e.g.][]{Heng2016AWAVELENGTHS,Stevenson2016QUANTIFYINGATMOSPHERES, Fu2017Planets,Crossfield2017TrendsExoplanets}. 

Opacity by clouds and hazes acts to mute the molecular absorption bands observed in the infrared, and characterisation of clouds and hazes is thus crucial to 
determining accurate abundances of molecules such as water and carbon monoxide 
\citep[e.g.][]{Benneke2012AtmosphericSpectroscopy, Griffith2014DisentanglingExoplanets, Line2013AHAT-P-12b, Sing2016ADepletion, Barstow2017ATransmission, Wakeford2017HAT-P-26b:Abundance, Heng2017TheChallenge, Wakeford2018TheConstraint, Pinhas2019H2OExoplanets}. Clouds and hazes are best characterised in the optical, where they are not degenerate with molecular abundances, and work with LRG-BEASTS and other surveys have shown that ground-based observations can provide optical transmission spectra with precision comparable to space-based telescopes \citep[e.g.][]{Todorov2019Ground-basedHAT-P-1b,Alderson2020LRG-BEASTS:WASP-21b,Carter2020DetectionWASP-6b,Weaver2021ACCESS:HAT-P-23b}. 
With the upcoming launch of the James Webb Space Telescope (JWST), which 
will provide precise transmission spectra across the near- and mid-infrared, 
the measurement of optical transmission spectra will be crucial for combined analyses to break degeneracies due to clouds and hazes and measure accurate atmospheric abundances. 

The Low Resolution Ground-Based Exoplanet Atmosphere Survey using Transmission Spectroscopy (LRG-BEASTS; ‘large beasts’) is a survey initiated in 2016 with the aim to gather a large sample of transmission spectra of exoplanet atmospheres, mainly focused on hot Jupiters at optical wavelengths. 
LRG-BEASTS has shown it is possible to achieve precisions on 4-m class telescopes that are comparable to those obtained with 8- to 10-m class telescopes and HST. Characterisations within LRG-BEASTS include the detection of hazes, Rayleigh scattering and grey clouds in atmospheres of the exoplanets WASP-52\,b \citep{Kirk2016TransmissionSurface,Louden2017AWASP-52b}, HAT-P-18\,b \citep{Kirk2017RayleighHAT-P-18b}, WASP-80\,b \citep{Kirk2018LRG-BEASTSWASP-80} and WASP-21\,b \citep{Alderson2020LRG-BEASTS:WASP-21b}, as well as sodium absorption in the atmosphere of WASP-21\,b \citep{Alderson2020LRG-BEASTS:WASP-21b}.  An analysis of the atmosphere of WASP-39\,b revealed supersolar metallicity \citep{Kirk2019LRG-BEASTS:WASP-39b}, and evidence for TiO was found in the atmosphere of the ultrahot Jupiter WASP-103\,b \citep{Kirk2021ACCESSWASP-103b}.

In this paper, we present the first 
transmission spectrum for the highly-inflated hot Jupiter \mystar\,b \citep{Neveu-Vanmalle2014WASP-94System}. 
This is 
also the first exoplanet transmission spectrum using long-slit spectroscopy with the EFOSC2 instrument at the New Technology Telescope (NTT). 

\mystar\,b  was detected in 2014 within the Wide Angle Search for Planets survey \citep[WASP,][]{Pollacco2006TheCameras}.
WASP-94 is a binary star system with each star hosting one known exoplanet. \mystar\ is a star of spectral type F8 with a transiting hot Jupiter orbiting with a period of $3.95$~days, 
while WASP-94B is a system consisting of a F9 type star and a non-transiting hot Jupiter with an orbital period of $2.008$~days characterised solely by radial velocity measurements so far \citep{Neveu-Vanmalle2014WASP-94System}. The stellar parameters for the WASP-94 system are summarised in Table~\ref{tab:wasp-94_stars}. GAIA DR2 determined the distance of the system to be $212.46 \pm 2.50$~pc \citep[GAIA DR2, ][]{Bailer-Jones20182}. 

\mystar\ and its companion star WASP-94B are of almost identical spectral type with F8 and F9 and of similar V magnitude with 10.1 and 10.5 respectively, and they have an angular separation of $15.03\pm 0.01$~arcseconds. This makes them excellent comparison stars for each other, which is highly favourable for ground-based transmission spectroscopy.  The two stars have a physical separation of at least $2700$~AU  \citep{Neveu-Vanmalle2014WASP-94System}.

\mystar\,b has an inflated radius with $1.72^{+0.06}_{-0.05}$~R$_{\textrm{Jup}}$ \citep{Neveu-Vanmalle2014WASP-94System} and a mass of $0.456^{+0.034}_{-0.036}$~M$_{\textrm{Jup}}$ \citep{Bonomo2017ThePlanets}. By measuring the Rossiter-McLaughlin effect, \citet{Neveu-Vanmalle2014WASP-94System} found that the orbit of \mystar\,b is misaligned and likely retrograde with a spin-orbit obliquity of $\lambda = 151^\circ \pm 20^\circ$. In addition, with its low density  and an equilibrium temperature of $1508 \pm 75$~K \citep{Garhart2020Eclipses}, one atmospheric scale height of \mystar\,b corresponds to a transit depth of 
262\,ppm, 
making it an 
attractive
target for atmospheric studies. No transmission spectrum has yet been published of \mystar\,b, making this the first atmospheric characterisation for this highly-inflated hot Jupiter.
The orbital and planetary parameters for \mystar\,b are summarised in Table~\ref{tab:wasp-94_planets}.

The structure of the paper is as follows. We describe our observations in Section~\ref{sec:observations}. This is followed by the reduction and analysis of our data in Section~\ref{sec:reduction} and Section~\ref{sec:light_curves}, respectively. In Section~\ref{sec:discussion} we discuss our results and in Section~\ref{sec:conclusions} we summarise our conclusions. 

\begin{table}

    \centering
    \caption{Stellar parameters for the binary star system WASP-94 as found by [1] \citet{Neveu-Vanmalle2014WASP-94System},  [2] \citet{Teske2016THEB}, [3] \citet{Bonomo2017ThePlanets} and [4] \citet[GAIA DR2, ][]{Bailer-Jones20182}.}
    \begin{tabular}{l c c c }
    \hline
        Parameter & WASP-94A &  WASP-94B & Reference\\ \hline
        $V_{\textrm{mag}}$ & 10.1 & 10.5 & [1]\\
        Spectral type & F8 & F9 & [1]\\
        $T_{\textrm{eff}}$ (K) & $6194 \pm 5$ & $6112 \pm 6$ & [2]\\
        Age (Gyr) & $2.55 \pm 0.25$ & $2.55 \pm 0.25$ & [2] \\
        log $g$ (log$_{10}$(cm/s$^{2}$)) & $4.210 \pm 0.011$ & $4.300 \pm 0.015$ & [2]\\
        $[$Fe/H$]$ & $0.320 \pm 0.004$ & $0.305 \pm 0.005$ & [2] \\
        Mass ($M_\odot$) & $1.450 \pm 0.090$ & $1.24 \pm 0.09$ & [3], [1]\\
        Radius ($R_\odot$) & $1.653^{+0.087}_{-0.081}$ & $1.438^{+0.067}_{-0.240}$ & [4]\\
        
    \hline
    \end{tabular}
    
    \label{tab:wasp-94_stars}
\end{table}

\begin{table}
\caption{Orbital and planetary parameters for the hot Jupiter \mystar\,b and the corresponding references.
[1] \citet{Neveu-Vanmalle2014WASP-94System}; [2] \citet{Bonomo2017ThePlanets}; [3] \citet{Garhart2020Eclipses} }
    \centering
    \begin{tabular}{l c c}
\hline
        Parameter & Value  & Reference\\ \hline
        Period, P (days) & $3.9501907^{+44}_{-30}$ &[1] \\
        Semi-major axis, a (AU) & $0.0554^{+0.0012}_{-0.0011}$ &[2] \\
        Mass, $M_\textrm{p}$ ($M_{\textrm{Jup}}$) & $0.456^{+0.034}_{-0.036}$ &[2] \\
        Radius, $R_\textrm{p}$ ($R_{\textrm{Jup}}$) & $1.72^{+0.06}_{-0.05}$ &[1] \\
        Inclination, i ($^\circ$) & $88.7 \pm 0.7$ &[1] \\
        Surface gravity, log g (log$_{10}$(cm/s$^{2}$)) & $2.590^{+0.044}_{-0.042}$ &[2] \\
        Equilibrium temperature, $T_\textrm{eq}$ (K) & $1508 \pm 75$ & [3] \\
        Spin-orbit obliquity $\lambda$ ($^\circ$) & $151 \pm 20$ & [1]
        \\
    \hline
    \end{tabular}
    
    \label{tab:wasp-94_planets}
\end{table}

\section{Observations}
\label{sec:observations}

\begin{figure*}
    \centering
    \includegraphics[width=\textwidth]{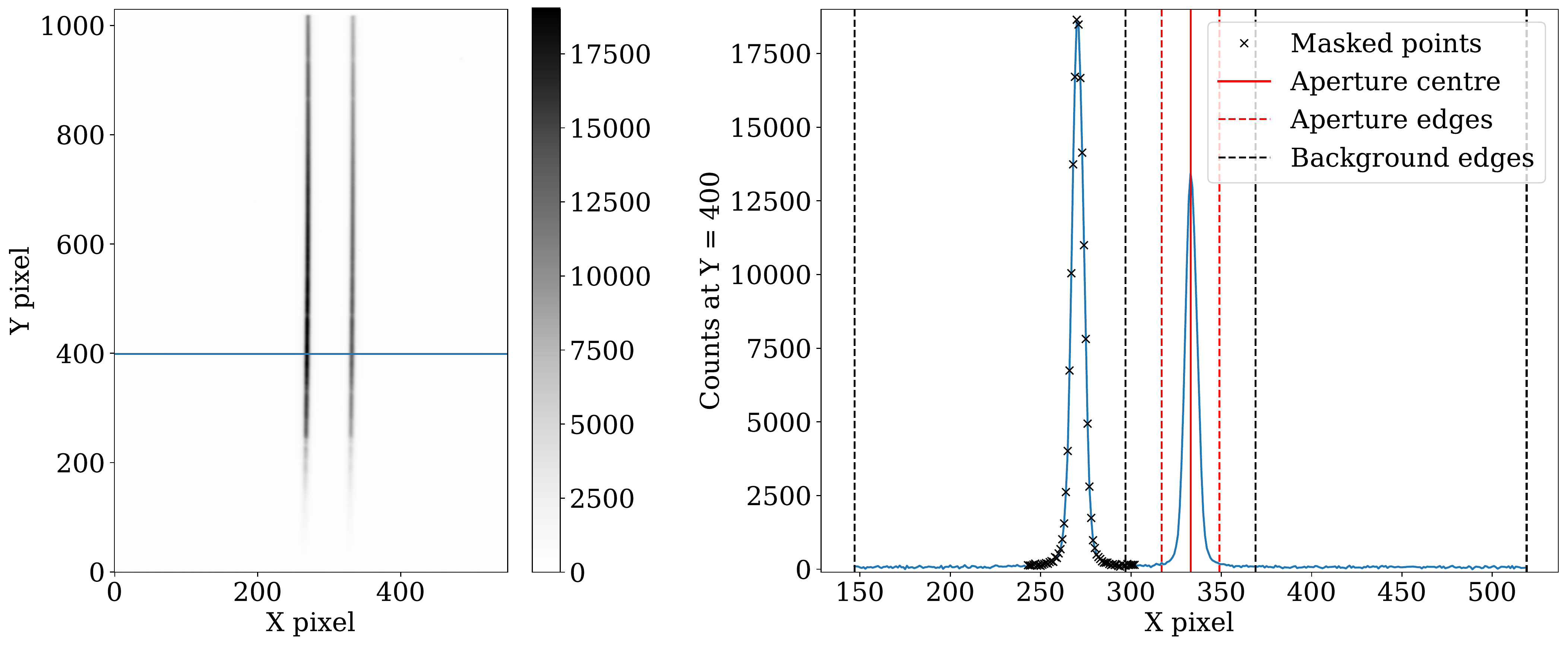}
    \caption{Left: Example science frame with the target \mystar\ (left) and comparison star WASP-94B (right). The horizontal and vertical axis correspond to the position along the slit and the wavelength, respectively. Note that this is a cropped frame (full frame in 2$\times$2 binning is 1030$\times$1030\,pixels). The ADU counts are shown with the colourbar. The blue line indicates the position of the cut corresponding to the righthand figure. \newline Right: A cut along spatial direction at row 400 showing the flux of the two stars in blue. In this example the right star was fitted with the red lines indicating the aperture: dashed lines representing the edges and the solid one the centre. The background regions are indicated with 
    dashed black lines, while the black crosses mark the mask for the non-fitted star. }
    \label{fig:example_frame_and_cut}
\end{figure*}

Observations of \mystar\ took place on the night of the 14\th\ of August 2017, using the EFOSC2 instrument \citep{Buzzoni1984TheEFOSC} mounted at the Nasmyth B focus of the ESO NTT, La Silla, Chile\footnote{Based on observations collected at the European Southern Observatory under ESO programme 099.C-0390(A).}. The detector is a Loral/Lesser CCD with a size of 2048 x 2048 pixels, a resolution of 0.12~arcseconds per pixel and an overall field of view of 4.1 arcmin. The fast readout mode was used and we applied $2\times2$ pixel binning, resulting in a readout time of 22\,s. 

For our spectroscopic measurements we chose a slit that was custom-built on our request, with a width of 27 arcseconds to avoid differential slit losses between target and comparison star. 
We also used
grism \#11, which provides a spectrum from $3750 - 7520$\,\AA\ at resolution of $R\sim 200$. 

We acquired 477 spectral frames with airmass ranging from 1.70 to 1.00 to 2.41. An exposure time of 30~s was used at the beginning of the night for 33~frames, afterwards it was changed to a 45~s exposure to increase the Signal to Noise Ratio (SNR). The moon was illuminated $48\%$ and rose during the second half of the night having a distance of at least $102^\circ$ to the target at all times. 39 bias frames were taken, as well as 
127
flat frames with different settings (91 dome flats, 25 sky flats, 11 lamp flats) and 15 HeAr arc frames for wavelength calibration, which were taken in the morning after the transit observations. Note that we did not use any of the flat frames during our final data reduction as using flat-fielding resulted in an increase in noise. This has been seen previously in the analysis of low-resolution spectra in LRG-BEASTS observations \citep[e.g.\ ][]{Kirk2017RayleighHAT-P-18b,Alderson2020LRG-BEASTS:WASP-21b,Kirk2021ACCESSWASP-103b}, and has also been found with ACCESS data \citep{Rackham2017ACCESSPHOTOSPHERE,Bixel2019ACCESS:WASP-4b,Espinoza2019ACCESS:Magellan/IMACS,Weaver2020ACCESS:K,Kirk2021ACCESSWASP-103b}.

WASP-94B served as a comparison star during the observations of \mystar\,b in order to perform differential spectrophotometry to reduce the effects of 
the Earth's atmosphere. \mystar\ and WASP-94B have similar spectral types and their angular separation of $15$~arcseconds \citep{Neveu-Vanmalle2014WASP-94System} is low enough to ensure very similar perturbations due to the Earth's atmosphere for both stars, but large enough so that their respective signals are uniquely identified (see  Fig.~\ref{fig:example_frame_and_cut}). This makes the system an ideal target for long-slit transmission spectroscopy.


\section{Data Reduction}
\label{sec:reduction}
The data were reduced using a custom-built \textsc{python} pipeline developed for the LRG-BEASTS survey and described in detail by \citet{Kirk2018LRG-BEASTSWASP-80} and used by \citet{Kirk2017RayleighHAT-P-18b, Kirk2019LRG-BEASTS:WASP-39b,Kirk2021ACCESSWASP-103b} and \citet{Alderson2020LRG-BEASTS:WASP-21b}. For this analysis we modified the methods for cosmic ray rejection and wavelength calibration, as described below. 

First, a master bias was constructed by median-combining 39 individual bias frames, and this was subtracted from each science frame. In contrast to previous LRG-BEASTS analyses, cosmic rays were not removed after the spectra were extracted, but at the raw science frame level. The change of method was chosen to increase the reliability and sensitivity of cosmic ray detection, and to minimise the risk of introducing spurious features into 1D spectra by mistakenly classifying real spectral features as cosmic rays. 

By dividing each science image by the previous frame (in the case of the first frame the subsequent frame was used) we identified affected pixels and  replaced them with the median of the surrounding pixels. For this we used a criterion of $16~\sigma$ with $\sigma$ being the standard deviation of the divided image, resulting in $\sim 50$ pixels being replaced per frame. The median of the surrounding pixels as a replacement value for outliers was chosen instead of the average as the latter can be biased towards a higher value as the rise in flux due to cosmic rays commonly spreads over several pixels.

To extract the spectra an aperture width of $33$\,pixels was used, as shown in Fig.\,\ref{fig:example_frame_and_cut}, and spectral counts were summed within the aperture. The sky background was fitted using a quadratic polynomial on a region of $150$~pixels on either side of the trace, separated by $20$~pixels from the edge of the source aperture. We experimented with several different values for aperture width, background width, offset and polynomial order for the background fit; the numbers stated here were found to be optimal for minimising the noise. Errors were calculated based on photon and read noise.

\begin{figure}
    \centering
    \includegraphics[width=\columnwidth]{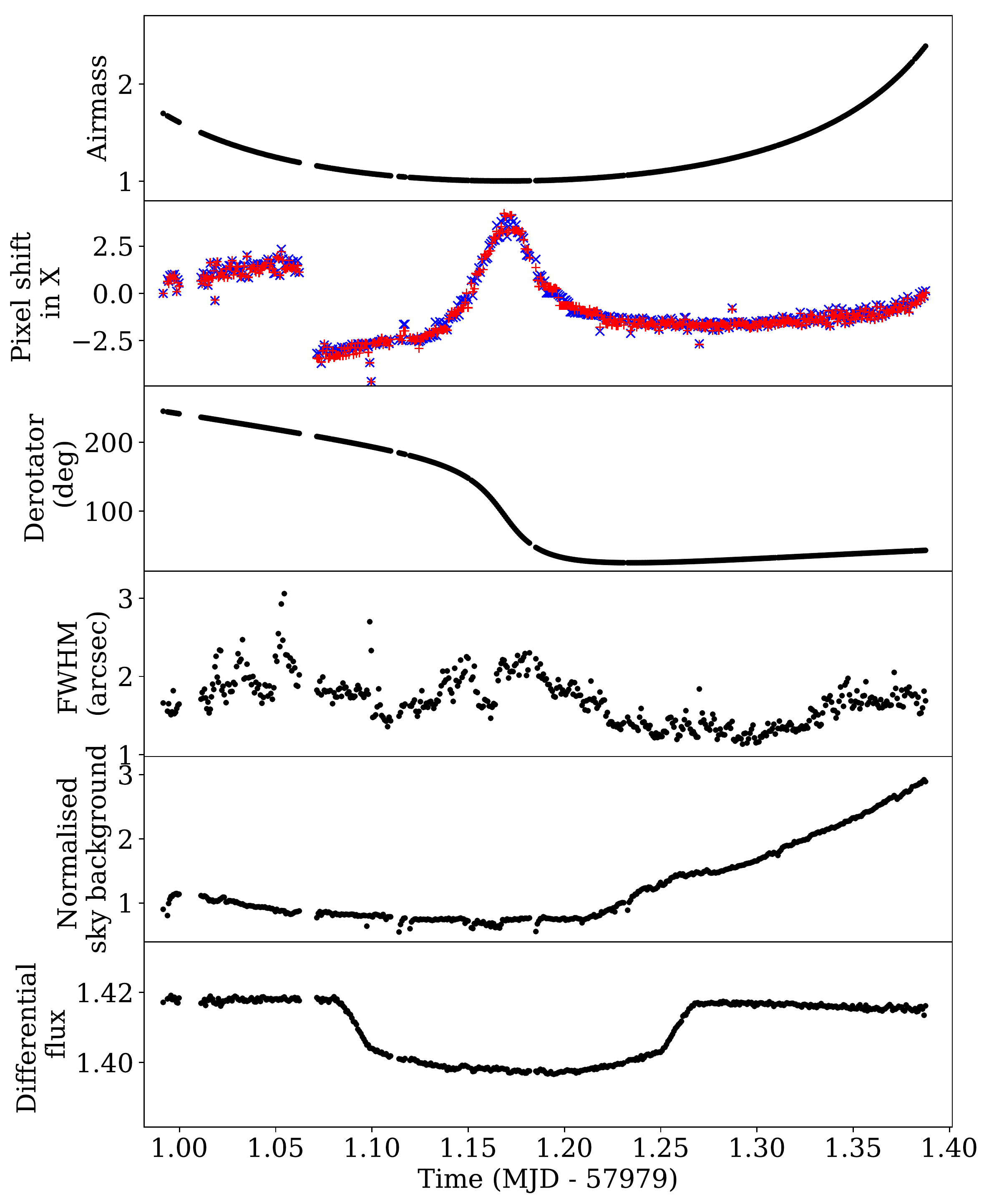}
    \caption{Ancillary data for the transit of \mystar\,b, with the colours blue and red corresponding to the target star \mystar\ and comparison star WASP-94B respectively. From top to bottom: airmass, pixel shift along the slit, derotator angle, full width half maximum (FWHM) of the stellar profile, normalised sky background and the differential flux (white light curve). }
    \label{fig:ancillary_plots}
\end{figure}

The target and comparison stars are $\sim 70$~pixels apart and thus would disturb the sky background fit, and so the star not being fitted had to be masked (see Fig.~\ref{fig:example_frame_and_cut}). Outliers of at least three standard deviations were also masked from the fit. Diagnostics of the spectral extraction and observing conditions are plotted in Fig.~\ref{fig:ancillary_plots}.

Following the extraction of each spectrum, wavelength solutions were found for each individual frame. Unlike in previous LRG-BEASTS studies, we used RASCAL \citep{Veitch-Michaelis2019RASCAL:Calibration} to make an initial wavelength calibration using the gathered HeAr arc frames. For improved accuracy throughout the time series, we combined the arc calibration  with a solution based on absorption lines in the extracted stellar spectra. Note that this wavelength calibration step was computed for each frame individually to account for wavelength drifts throughout the night (on the order of $\sim 5$\,pixels or $\sim 25$\,\AA). 

\begin{figure}
    \centering
    \includegraphics[width=\columnwidth]{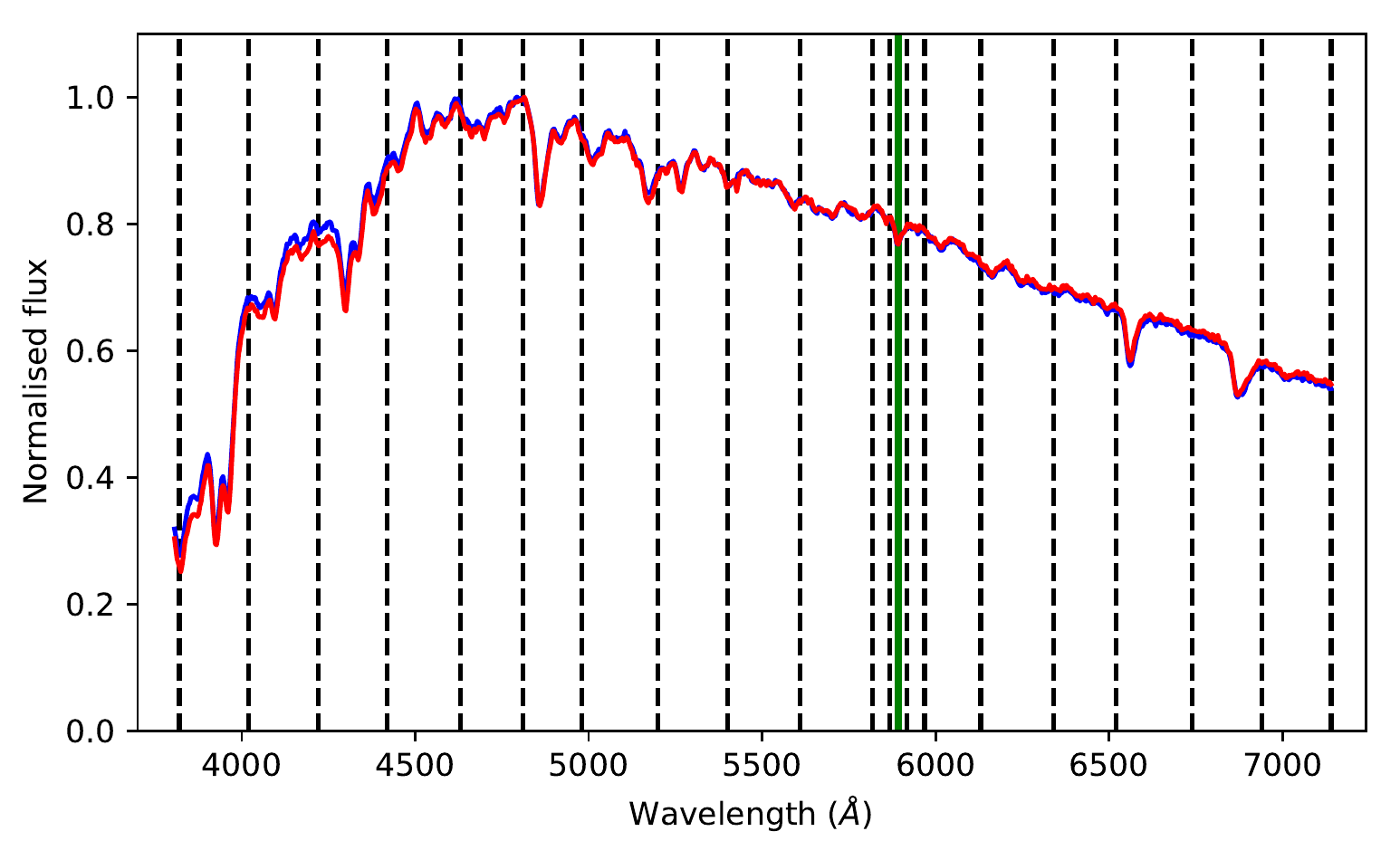}
    \caption{The blue and red line correspond to the flux of \mystar \ and the comparison star. The dashed black lines indicate the edges of the wavelength bins, the green line shows the central wavelength of the sodium doublet line.  }
    \label{fig:wavelength_bins}
\end{figure}

For our main analysis, the spectra were binned into 19 wavelength bins. Of these, 16 have a width in the range $160-220$\,\AA, with the wavelength ranges chosen to avoid bin edges falling on spectral absorption lines.  The remaining three bins have a width of $50$\,\AA\ centred on the sodium doublet. The wavelength bins are shown in Fig.~\ref{fig:wavelength_bins}. We also extracted higher-resolution data around the sodium doublet for analysis of the line width in Section\,\ref{sec:line_width} (35 bins with a width of 14\,\AA). The light curves for each bin were computed by summing the flux within the corresponding wavelength range of each frame. To correct for the affects of the Earth's atmosphere each light curve was divided by the comparison star's light curve. A white-light light curve was also computed 
by defining the whole spectrum as a single wavelength bin. 

\section{Data Analysis}
\label{sec:light_curves}
\subsection{Transit model}

\begin{table*}
    \centering
    \caption{Parameter values obtained from the white-light curve fitting, for both the polynomial and Gaussian Processes detrending. The respective priors placed on these parameters are also listed.  Values for semi-major axis $a$, radius of the star $R_*$ and radius of the planet $R_p$ and inclination $i$ are listed in Table~\ref{tab:wasp-94_planets}. The values for the parameters for $a/R_*$, $i$ and $T_C$ listed here were fixed for the spectroscopic light curve fitting. }
    \label{tab:system_parameters}
    \begin{tabular}{lccccc}
    \hline
    Parameter &  \multicolumn{2}{c}{Prior distribution and range} & Polynomial detrending & GP detrending\\ \hline
   Scaled stellar radius $a/R_*$ & Uniform & $a/R_* \pm  3\sigma_{a/R_*}$ &  $7.368^{+0.028}_{-0.021}$  & $7.436^{+0.020}_{-0.029}$\\
   Inclination $i$ ($^\circ$) & Uniform & $i$ $\pm$ 3$\sigma_{i}$ & $89.25^{+0.20}_{-0.37}$ & $89.42 \pm 0.38$\\
    Time of mid-transit $T_C$ (BJD) & Uniform & $0.9 \times\ T_C$, $1.1 \times\ T_C$ & $2457980.681061 \pm 0.000038$  &  $2457980.68093 \pm 0.00019$\\
    Transit depth $R_p/R_*$  & Uniform & $R_p/R_*$ $\pm$ 3$\sigma_{R_p/R_*}$& $0.10827\pm 0.00017$  & $0.10544^{+ 0.00040}_{-0.00039}$\\
    Limb-darkening coefficient $u1$ & Uniform & $u1 \pm 4\sigma_{u1}$ & $0.5326\pm 0.0049$ & $0.529^{+0.021}_{-0.018}$\\
    Limb-darkening coefficient $u2$ & Fixed & -- & 0.0862 & 0.0862\\
    \hline
    \end{tabular} 
\end{table*}

In order to fit the individual transit light curves we used the nested sampling algorithm \textsc{PolyChord} \citep{Handley2015PolyChord:Cosmology,Handley2015PolyChord:Sampling} in \textsc{python} in combination with the \textsc{batman} package \citep{Kreidberg2015BatmanPython} and the analytic light curves from \citet{Mandel2002AnalyticSearches}.

The fitting parameters included the ratio of planet to star radius, $R_p/R_*$, the inclination $i$ of the system, the quadratic limb-darkening coefficients $u1$ and $u2$, the scaled stellar radius $a/R_*$ and the time of mid-transit $T_C$. Additional parameters were used to describe detrending functions, as explained below. 

First, the system parameters $a/R_*$, $i$ and $T_C$ were fitted for using the white-light light curve (see Table~\ref{tab:system_parameters}). All priors for system parameters were chosen to be wide and uniform, with lower and upper limits corresponding to the respective literature values (Table~\ref{tab:wasp-94_planets}) subtracting and adding three times the respective error. The limb-darkening coefficients and respective uncertainties were generated using the Limb-Darkening Toolkit (\textsc{LDTk}) package \citep{Parviainen2015LDTK:Toolkit}. A quadratic limb-darkening law was applied and one coefficient held fixed to avoid degeneracies ($u2$). The prior for $u1$ was uniform, centered at the generated value with a range corresponding to the generated error inflated by a factor of four. All priors are summarised in Table~\ref{tab:system_parameters}.

The parameter values determined from the white-light light curve 
were then used as fixed parameters for the binned light curve fits to ensure the errors in the transmission spectrum reflected the uncertainties in the wavelength-dependence of the planet radius, and not the uncertainty in the absolute radius.

    

To model 
systematic trends
in the light curves we took two different approaches for detrending in order to demonstrate that our transmission spectrum is not sensitive to the treatment of systematics. 

\subsection{Polynomial Detrending}
\label{sec:poly}
\begin{figure}
\centering
    \subfigure[Polynomial]{\includegraphics[width=\columnwidth]{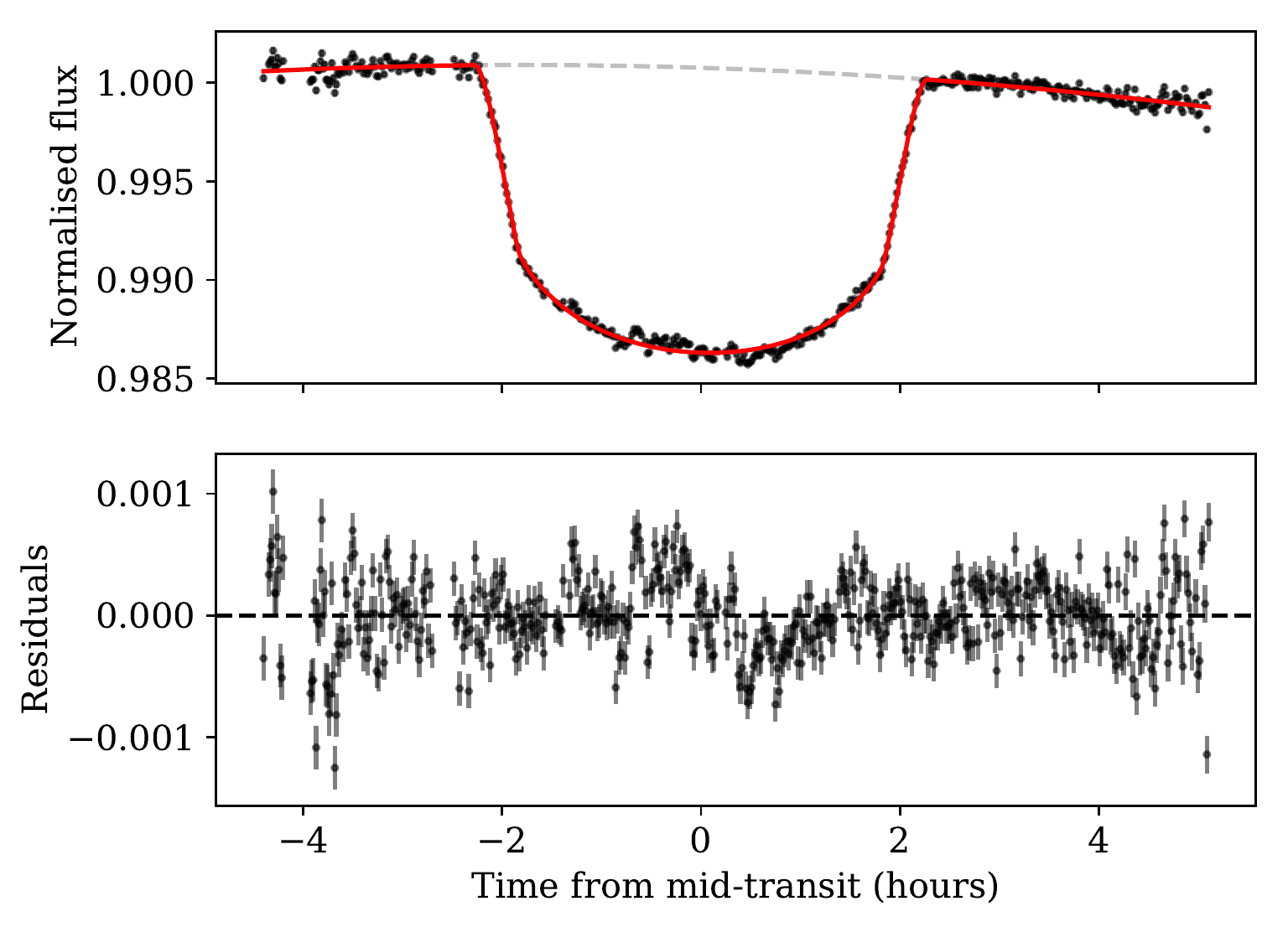}}
    \subfigure[Gaussian Process]{\includegraphics[width=\columnwidth]{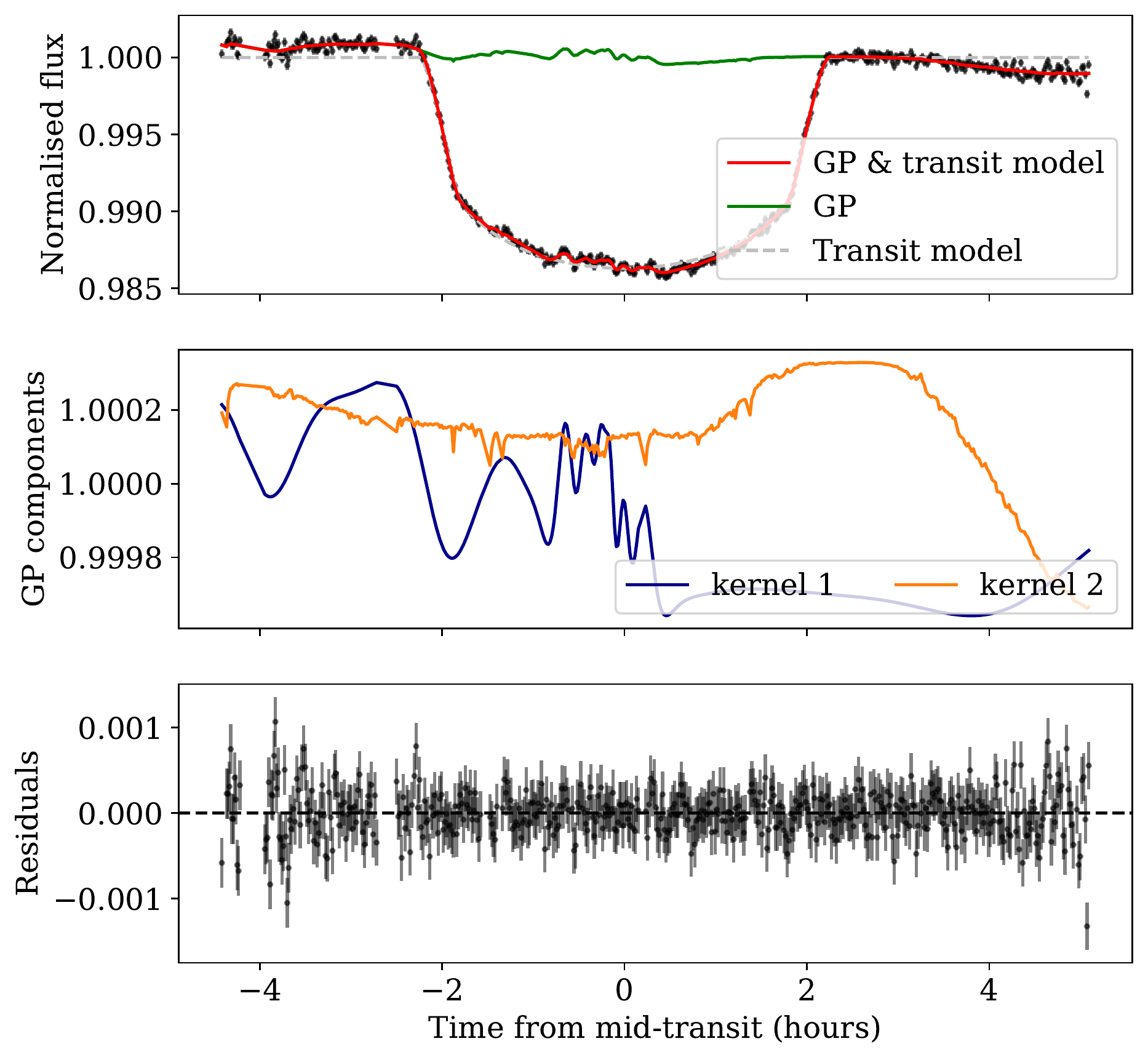}}
    \caption{Our white-light transit light curve of \mystar\,b fitted using (a) a quadratic in time detrending function and (b) a GP. The respective top panels show the fitted light curve; bottom panels show residuals to the fits.  In both (a) and (b) the red line shows the best fitting model, while the dashed grey line represents the transit model including the quadratic detrending model in (a) and solely the transit model without detrending in (b). In (b) the green line refers to the overall GP model, and the orange and blue lines show the contributions from the two GP kernel inputs: sky and derotator angle respectively.}
    \label{fig:nonGP_WL}
\end{figure}

\begin{figure*}
    \centering
    \includegraphics[width=\textwidth]{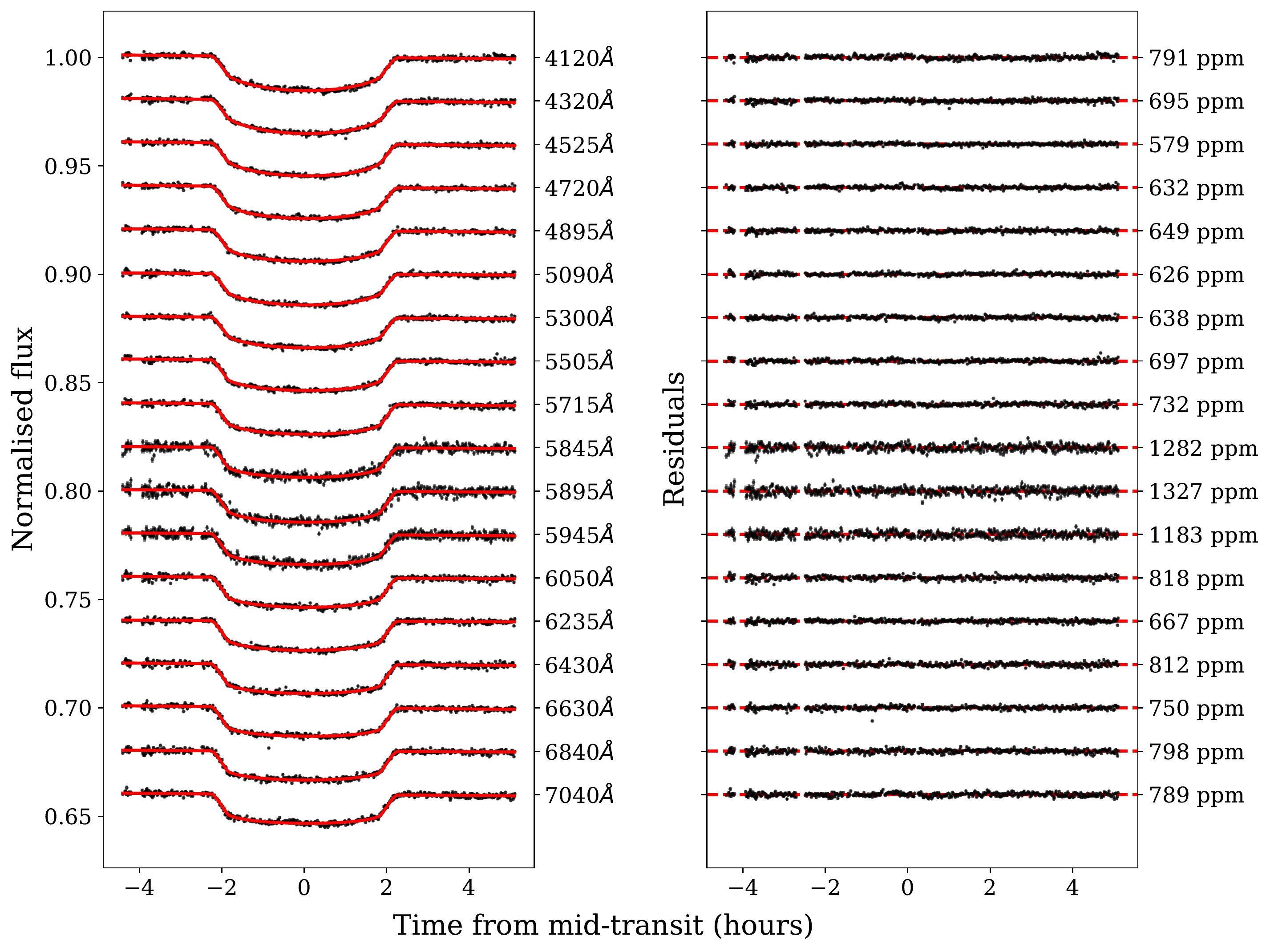}
    \caption{Left: Light curves for each individual wavelength bin fitted with a transit model and a linear in time detrending model. The wavelength increases from blue to red from top to bottom. The red lines correspond to the fitted transit models including the linear trend. Right: Respective RMS amplitude of the residuals for the individual fitted spectroscopic light curves on the left.   }
    \label{fig:nonGP_all_lightcurves}
\end{figure*}

A low-order polynomial model was fitted in combination with the transit model to remove a small overall trend (see Figs.~\ref{fig:nonGP_WL}\,\&\,\ref{fig:nonGP_all_lightcurves}). The best fitting model for the white-light light curve was quadratic in time (Fig.~\ref{fig:nonGP_WL}), while the light curves for the individual wavelength bins were found to be fitted best by a linear model (Fig.~\ref{fig:nonGP_all_lightcurves}).

We experimented with several variations of polynomial functions and different input files e.g.\ polynomials of different order with airmass, sky background, telescope derotator angle or two or more linear in time and airmass models simultaneously. However, the linear polynomial in time showed the highest Bayesian evidences with strong statistical significance i.e.\ at least a difference in logarithmic evidence of $>5$ corresponding to a 99.3~$\%$ probability \citep{Jeffreys1983TheoryProbability}. 

The system parameters $a/R_*$, $i$ and $T_C$ obtained from the white-light light curve are shown in Table~\ref{tab:system_parameters}, while the parameters $R_p/R_*$, $u1$, $u2$ for each individual wavelength bin are shown in Table~\ref{tab:nonGP_transmission_spectrum}.

\begin{table}
    \centering
    \caption{Transmission spectrum of \mystar\,b in tabulated form using a linear in time detrending, as plotted in Fig.~\ref{fig:transmission_spectrum_both} }
    \label{tab:nonGP_transmission_spectrum}
        \begin{tabular}{cccc}
    \hline
        Bins ($\si{\angstrom}$)  & $R_p/R_*$ & u1 & u2  \\ \hline
         3800 - 4020 & $0.10790^{+0.00045}_{-0.00067}$ & $0.87^{+0.02}_{-0.03}$ & -0.0639\\
         4020 - 4220  & $0.10965^{+0.00036}_{-0.00061}$ & $0.70^{+0.03}_{-0.02}$ & -0.0109\\
         4220 - 4420 & $0.10831^{+0.00033}_{-0.00064}$ & $0.72^{+0.03}_{-0.02}$ & -0.0275\\
         4420 - 4630  & $0.10774^{+0.00027}_{-0.00059}$ & $0.64^{+0.03}_{-0.01}$ & 0.0460\\
         4630 - 4810  &  $0.10766^{+ 0.00025}_{- 0.00070}$ & $0.58^{+0.03}_{-0.01}$ & 0.0736\\
         4810 - 4980  & $0.10740^{+ 0.00031}_{- 0.00067}$ &  $0.53^{+0.03}_{-0.02}$ & 0.1032\\
        4980 - 5200  & $0.10709^{+0.00028}_{-0.00065}$ &  $0.55^{+0.03}_{-0.01}$ &  0.0884\\
       5200 - 5400 & $0.10636^{+0.00029}_{-0.00072}$&  $0.51^{+0.03}_{-0.01}$ & 0.1080\\
        5400 - 5610  & $0.10724^{+0.00023}_{-0.00079}$ &  $0.46^{+0.05}_{-0.01}$ & 0.1134\\
        5610 - 5818  & $0.10734^{+0.00025}_{-0.00076}$  & $0.45^{+0.04}_{-0.01}$  &  0.1265\\
        5818 - 5868 & $0.10757^{+0.00033}_{-0.0015}$  & $0.41^{+0.08}_{-0.01}$  & 0.1348\\
       5868 - 5918 & $0.10975^{+0.00038}_{-0.0015}$ &  $0.43^{+0.07}_{-0.01}$  &  0.1269\\
        5918 - 5968 & $0.10810^{+ 0.00037}_{- 0.0018}$  & $0.42^{+0.08}_{-0.01}$  &  0.1315\\
        5968 - 6130  & $0.10741^{+ 0.00021}_{- 0.0011}$  & $0.41^{+0.06}_{-0.01}$  &  0.1334\\
        6130 - 6340  & $0.10700^{+ 0.00025}_{- 0.00092}$ & $0.41^{+0.05}_{-0.01}$  & 0.1337 \\
        6340 - 6520  & $0.10745^{+ 0.00016}_{- 0.0011}$  & $0.36^{+0.06}_{-0.01} $ &  0.1404\\
        6520 - 6740  & $0.10725^{+ 0.00017}_{- 0.0011}$  & $0.32^{+0.07}_{-0.01} $ &  0.1565\\
        6740 - 6940 &  $0.10710^{+ 0.00030}_{- 0.00091} $ & $0.35^{+0.06}_{-0.01}$ & 0.1429 \\
        6940 - 7140&  $0.10710^{+ 0.00029}_{- 0.00093} $ & $0.35^{+0.06}_{-0.01}$ & 0.1434\\
\hline
    \end{tabular}
    
\end{table}

The first wavelength bin showed features in the light curve that are distinct from  the other bins, potentially related to increased extinction in the Earth's atmosphere and the much lower flux in this bin (see Fig.~\ref{fig:wavelength_bins}). While experimenting with different detrending methods as well as bin sizes we found that the transit depth for this particular bin changed significantly with the type of detrending used. Because of this poorly defined behaviour we conducted further analysis without considering this first wavelength bin and it is excluded from the white light fit, as well as Figs.\,\ref{fig:nonGP_all_lightcurves}\,\&\,\ref{fig:all_fitted_lightcurves_GP}.

\subsection{Detrending with Gaussian Process Regression}
\label{sec:gp}
Using Gaussian Processes (GPs) is a common technique in exoplanet research to model correlated noise \citep{Rasmussen2006GaussianLearning}. It has been used in detrending photometric data \citep[e.g.][]{Carter2009ParameterCurves,Csizmadia2015TransitingStar, Aigrain2016K2SC:Regression, Pepper2017183, Luger2017ATRAPPIST-1,Santerne2018AnComposition,Lienhard2020GlobalSurvey,Nowak2020The3780,Leleu2021SixTOI-178}, in disentangling planetary signals from stellar activity in radial velocity signals \citep[e.g.][]{Haywood2014PlanetsSystem,Rajpaul2015AData,Rajpaul2016GhostB,Faria2016UncoveringVelocities,SuarezMascareno2018The176986,Barragan2019RadialNeptune, Damasso2020TheThere,Ahrer2021TheDetection} and in transmission spectroscopy \citep[e.g.][]{Gibson2012ASpectroscopy,Gibson2012ProbingSpectroscopy,Gibson2013ANM,Gibson2013TheFeatures,Evans2015AModels,Hirano2016DWARF,Evans2016DetectionAtmosphere, Cartier20163,Louden2017AWASP-52b,Kirk2017RayleighHAT-P-18b,Kirk2018LRG-BEASTSWASP-80,Kirk2019LRG-BEASTS:WASP-39b,Kirk2021ACCESSWASP-103b}. 

For detrending our light curves with GP regression, we used the \textsc{george} package \citep{Ambikasaran2014FastProcesses} in combination with \textsc{PolyChord}. GPs use hyperparameters which describe functions modelling the covariance between the data points. In this case, the hyperparameters take the form of a length scale and an amplitude of the noise correlation. Note that we did not apply a common noise model or any other detrending method before fitting the spectroscopic light curves. 

For our fitting we used a combined kernel, a sum of the following basic kernels: a white noise kernel capturing random Gaussian noise and two squared-exponential kernels, one with sky background as an input variable, and the other using telescope derotator angle (see Figure~\ref{fig:ancillary_plots}). We want to highlight here that systematics based on the derotator are not expected to be very large for the NTT as there is no Atmospheric Dispersion Corrector (ADC), which has been known to cause large rotation-induced systematics in some VLT/FORS data \citep{Moehler2010CorrectionData}.


As with previous LRG-BEASTS papers, each input for the GP was standardised by subtracting the mean and dividing by the standard deviation, as previously also used by e.g.\ \citet{Evans2017AnStratosphereb}. We experimented with different combinations of kernel inputs which included up to five (airmass, FWHM, position of stars along the slit, telescope derotator angle, mean sky background). We found that two kernel inputs (sky and derotator angle) were sufficient to remove the red noise in the data (Fig.\,\ref{fig:nonGP_WL}). The use of additional kernel inputs resulted in a lower Bayesian evidence value ($\Delta \log \mathcal{Z}$ > 2), it did not improve the noise in the residuals and also led to degeneracies between kernel components. For this reason, we used two kernel inputs (sky and derotator angle) for our final analysis. 

Our white-light light curve fit is shown in the middle panel of Fig.\,\ref{fig:nonGP_WL}\, (b) where the GP component contributions of the sky and derotator as inputs are shown in orange and blue, respectively. The distributions of the posteriors of the white-light light curve fit including the GP hyperparameters are attached in the Appendix, Fig.\,\ref{fig:corner_GP_WL}. 

The final GP kernel function $k$ is then described as the sum of two exponential-squared basic kernel functions and a white noise kernel function:


\begin{equation}
\begin{split}
    k_{sd}(s_i,d_i,s_j,d_j) & = A\left(\exp \left(-\eta_s (s_i-s_j)^2 \right) + \exp \left(-\eta_d (d_i-d_j)^2 \right)\right) \\ & + \sigma^2 \delta_{ij},
\end{split}
\end{equation}

where $A$ is the amplitude of the covariance, $s$ and $d$ are the sky and derotator inputs, $\eta_s$ and $\eta_d$ are the sky and derotator inverse length scales, $\sigma^2$ is the variance of the white noise, and $\delta$ is the Kronecker delta. In practice, we fit for the natural logarithm of $A$, $\eta_s$, $\eta_d$, and $\sigma^2$ since these vary over orders of magnitude.


In contrast to previous LRG-BEASTS analyses, all spectroscopic bins were fit simultaneously, sharing the GP hyperparameters for length scale to capture the common noise shape, while allowing the amplitude and the white noise to be independently fitted for each spectroscopic light curve. This was computationally expensive, but resulted in more consistent detrending between neighbouring wavelength bins. The two shared parameters across all 18 bins are the natural logarithm of the inverse length scale of the kernel inputs sky and derotator angle $\ln \eta_s$ and $\ln \eta_d$.
Each spectroscopic light curve $j$ had four individual parameters: transit depth ($R_p/R_*$)$_j$, limb darkening coefficient $u1_j$, logarithm of white noise variance $\ln \sigma^2_j$ and GP amplitude $A_j$. In total 74 parameters were fitted simultaneously. 

\begin{figure*}
    \centering
    \includegraphics[width=\textwidth]{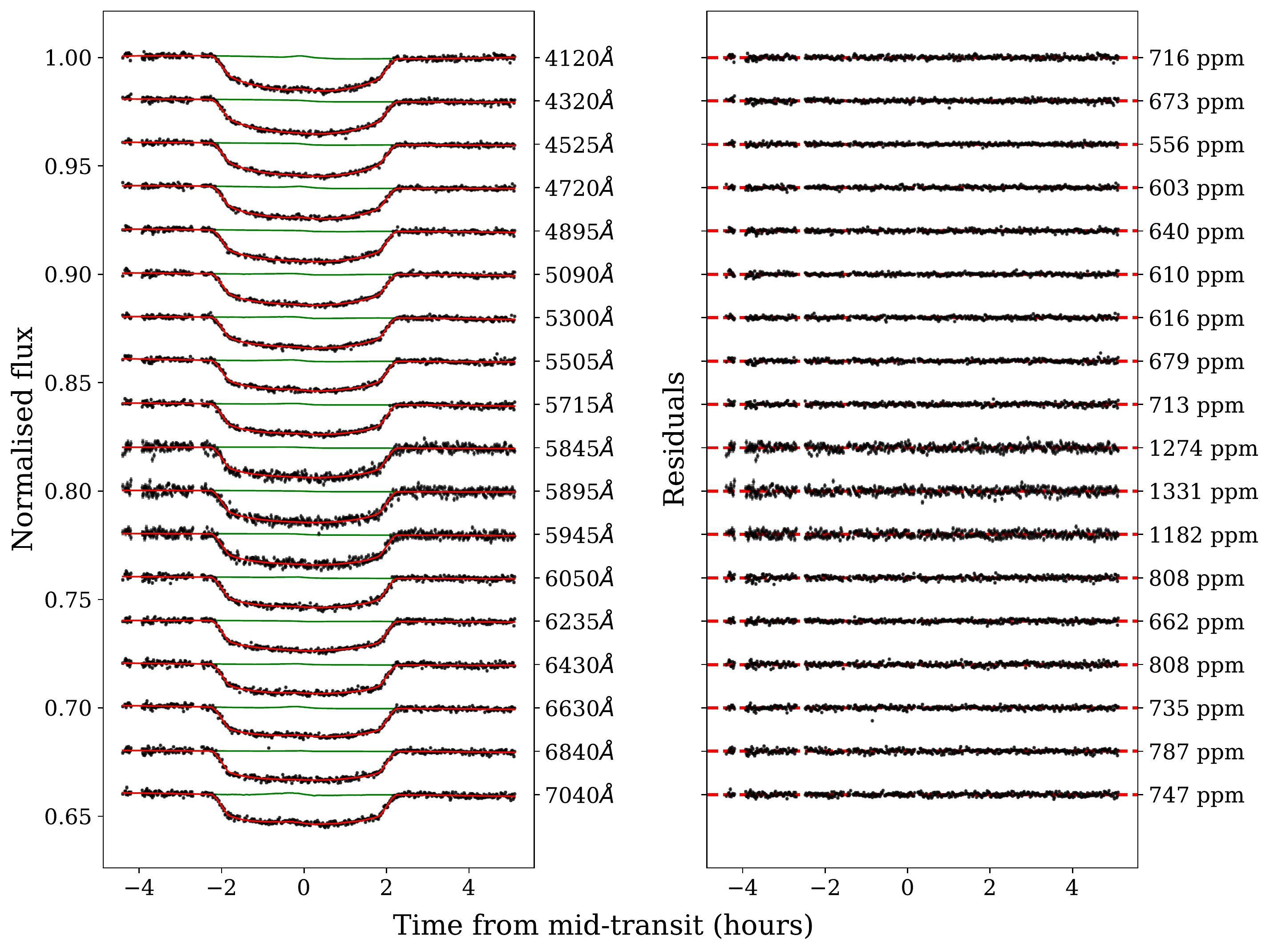}
    \caption{Left: Light curves for each individual wavelength bin 
    fitted with the transit model and the  GP detrending model (red line). The GP detrending model is also plotted on its own in green. Wavelength increases from top to bottom. Right: Corresponding Residuals for the spectroscopic light curves.  }
    \label{fig:all_fitted_lightcurves_GP}
\end{figure*}

The fitted spectroscopic light curves are shown in Fig.~\ref{fig:all_fitted_lightcurves_GP} and the resulting transmission spectrum in tabular form is displayed in Table~\ref{tab:GP_transmission_spectrum}, including the retrieved GP hyperparameters.

\begin{table}
    \centering
    \caption{Retrieved transmission spectrum of \mystar\,b in tabulated form using GP detrending, as plotted in Fig.~\ref{fig:transmission_spectrum_both}, and the individual retrieved GP hyperparameters amplitude $A$ and the white noise kernel variance parameter $\ln \sigma^2$. The shared hyperparameters are determined to be $\ln \eta_s = -1.21^{+0.90}_{-1.9}$ and $\ln \eta_d = 0.26^{+0.44}_{-0.58}$. The fixed values for the limb-darkening parameter $u2$ are equal to the ones used in the linear detrending and displayed in Table\,\ref{tab:nonGP_transmission_spectrum}. }
    \label{tab:GP_transmission_spectrum}
    \begin{tabular}{ccccc}
    \hline
        Bins ($\si{\angstrom}$)  & $R_p/R_*$ & u1 & ln A & $\ln \sigma^2$ \\ \hline
         4020 - 4220  & $0.10825^{+0.00062}_{-0.00067}$ & $0.71 \pm 0.03$  & $-14.1^{+3.5}_{-2.1}$ & $-17.0^{+1.6}_{-6.8}$\\
         4220 - 4420 & $0.10728^{+0.00062}_{-0.00066}$ & $0.73 \pm 0.03$  & $-14.9^{+3.7}_{-2.1}$ & $-18.4^{+2.6}_{-6.2}$\\
         4420 - 4630  & $0.10656^{+0.00057}_{-0.00057}$ & $0.65 \pm 0.03$  & $-15.3^{+3.5}_{-1.8}$ & $-21.0^{+3.9}_{-4.6}$\\
         4630 - 4810  &  $0.10630^{+0.00058}_{-0.00058}$ & $0.59 \pm 0.03$  & $-14.8^{+3.5}_{-2.0}$ & $-20.4^{+3.8}_{-5.0}$\\
         4810 - 4980  & $0.10629^{+ 0.00058}_{- 0.00060}$ &  $0.53 \pm 0.03$  & $-15.4^{+3.5}_{-2.1}$ & $-20.3^{+3.8}_{-5.0}$\\
        4980 - 5200  & $0.10654^{+0.00056}_{-0.00058}$ &  $0.55\pm 0.03$ & $-15.5^{+3.8}_{-2.4}$ & $-19.3^{+3.1}_{-5.7}$\\
       5200 - 5400 & $0.10618^{+0.00056}_{-0.00056}$&  $0.51 \pm 0.03$  & $-15.4^{+3.7}_{-2.3}$ & $-19.6^{+3.4}_{-5.4}$\\
        5400 - 5610  & $0.10590^{+0.00054}_{-0.00050}$ &  $0.48 \pm 0.03$ & $-14.8^{+3.6}_{-2.3}$ & $-16.8^{+1.3}_{-6.5}$\\
        5610 - 5818  & $0.10681^{+0.00058}_{-0.00061}$  & $0.46\pm 0.03$   & $-15.4^{+3.7}_{-2.3}$ & $-16.0^{+0.8}_{-5.5}$\\
        5818 - 5868 & $0.10662^{+0.00096}_{-0.00088}$  & $0.45\pm 0.04$   & $-17.7^{+4.2}_{-8.6}$ & $-20.0^{+4.6}_{-5.2}$\\
       5868 - 5918 & $0.10859^{+0.00070}_{-0.0011}$ &  $0.48 \pm 0.04$   & $-17.7^{+4.0}_{-8.1}$ & $-19.0^{+4.2}_{-5.9}$\\
        5918 - 5968 & $0.1073^{+ 0.0010}_{- 0.0011}$  & $0.46\pm 0.04$  & $-16.8^{+3.7}_{-6.5}$ & $-20.9^{+4.6}_{-4.6}$\\
        5968 - 6130  & $0.10668^{+ 0.00066}_{- 0.00068}$  & $0.43\pm 0.03$  & $-16.0^{+3.7}_{-2.5}$ & $-17.0^{+1.8}_{-6.9}$\\
        6130 - 6340  & $0.10637^{+ 0.00058}_{- 0.00059}$ & $0.42\pm 0.02$  & $-17.3^{+3.5}_{-2.6}$  & $-20.6^{+4.0}_{-4.8}$\\
        6340 - 6520  & $0.10647^{+ 0.00063}_{- 0.00065}$  & $0.40\pm 0.03 $  & $-15.9^{+3.7}_{-2.7}$ & $-16.6^{+1.5}_{-6.9}$\\
        6520 - 6740  & $0.10603^{+ 0.00061}_{- 0.00057}$  & $0.36\pm 0.03 $  & $-14.7^{+3.6}_{-2.1}$ & $-17.6^{+2.1}_{-6.7}$\\
        6740 - 6940 &  $0.10642^{+ 0.00063}_{- 0.00064} $ & $0.38\pm 0.03$  & $-17.0^{+3.8}_{-6.0}$ & $-19.1^{+3.4}_{-5.8}$\\
        6940 - 7140&  $0.10649^{+ 0.00069}_{- 0.00070} $ & $0.38\pm 0.03$  & $-14.5^{+3.7}_{-2.4}$ & $-21.0^{+4.2}_{-4.6}$\\
\hline
    \end{tabular}
    
\end{table}

\subsection{Transmission Spectrum}
\label{sec:Nyquist}
\begin{figure}
    \centering
    \includegraphics[width=\columnwidth]{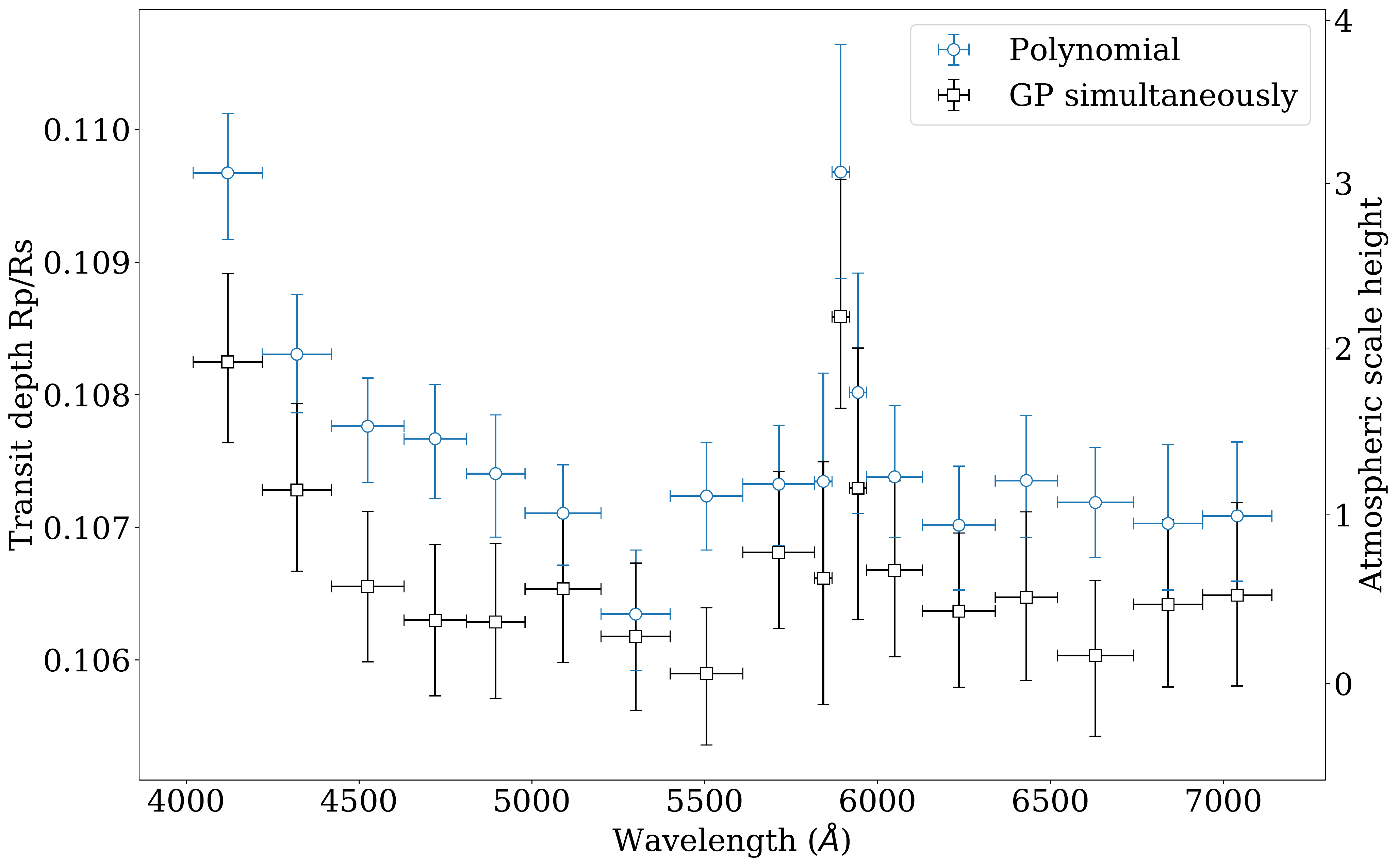}
    \caption{Transmission spectrum for \mystar\,b determined using two detrending models: a linear trend in time (blue), and using Gaussian Process with kernel inputs based on sky background and telescope derotator angle (black). Note that the offset between the transmission spectra is a natural consequence of fixing the values of system  parameters, which is necessary in order to accurately assess the relative uncertainties between different wavelengths. 
    }
    \label{fig:transmission_spectrum_both}
\end{figure}

Following the light curve fitting, the transmission spectra for both detrending methods were constructed as shown in Fig.~\ref{fig:transmission_spectrum_both}. While the transmission spectrum inferred with parametric detrending shows less uncertainty, the GP noise modelling 
provided a slightly better fit, 
showing less residual noise (see Figs.\,\ref{fig:nonGP_all_lightcurves}\,\&\,\ref{fig:all_fitted_lightcurves_GP}). The overall shape of both transmission spectra are very similar, with a slope towards the blue end of the spectrum and excess absorption at $5893$\,\AA\ (the centre of the NaI doublet). 
The 
detection of the sodium absorption feature in both spectra demonstrates that the presence of the signal is not sensitive to our treatment of the systematics.


\subsection{Testing the sodium signal}
\begin{figure}
    \centering
    \includegraphics[width=\columnwidth]{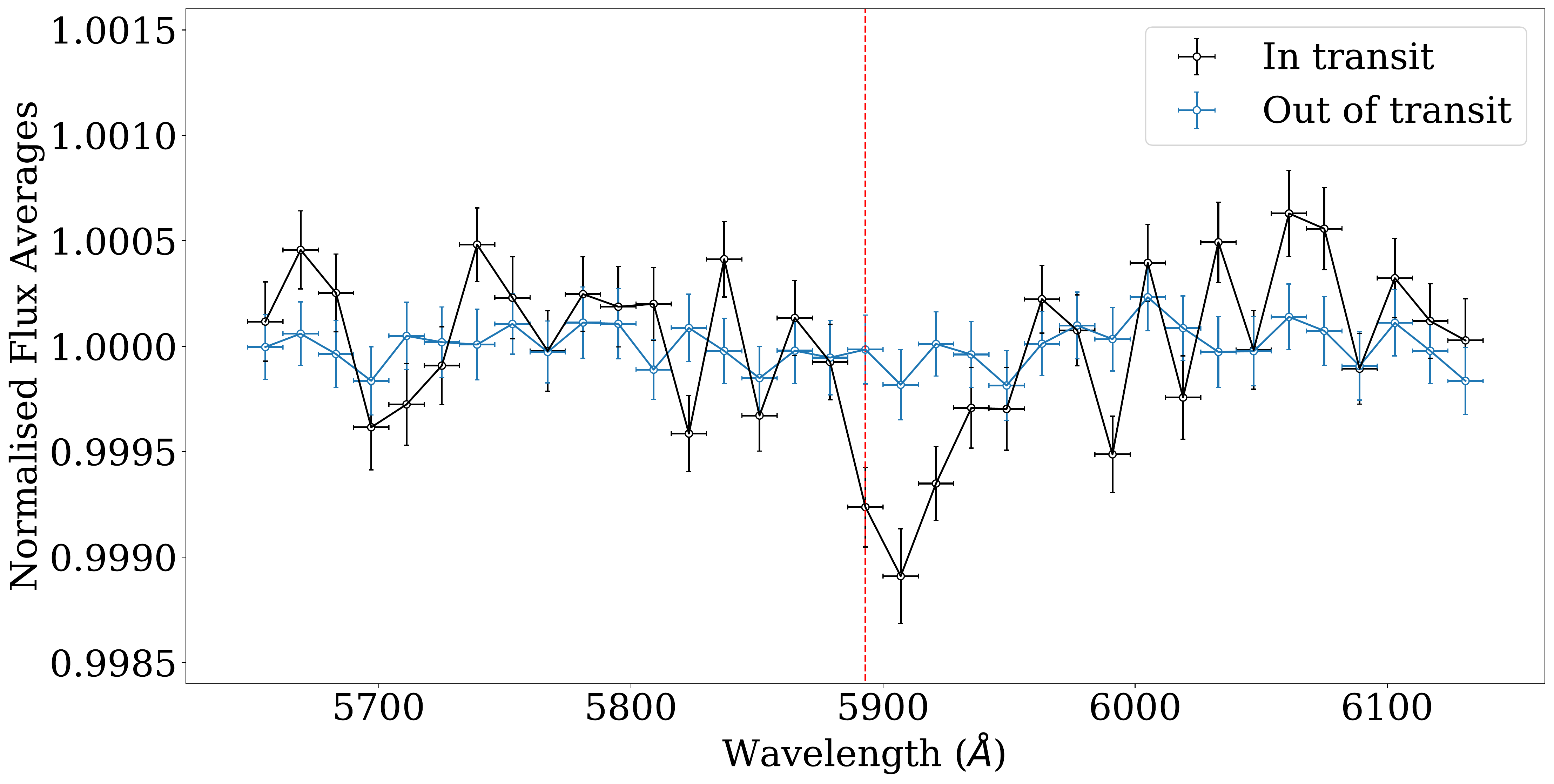}
    \caption{Direct ratio of the spectra of our target star (\mystar) and comparison star (WASP-94B) made separately for in-transit (black) and out-of-transit (blue) times. This shows excess absorption during transit at the wavelength of the sodium doublet (dashed red line).
    }
    \label{fig:Na_feature_flux_average}
\end{figure}

As an independent test of the significance of the sodium absorption, we extracted average spectra for \mystar\ and WASP-94B for in-transit (between 2nd and 3rd contact) and out-of-transit times, and calculated the direct ratio of the two spectra. These are plotted in Fig.\,\ref{fig:Na_feature_flux_average}. Because the two stars have such similar spectral types, stellar absorption features cancel, and the out-of-transit ratio spectrum is featureless (blue points in Fig.\,\ref{fig:Na_feature_flux_average}). In contrast, a strong absorption feature is seen at the expected wavelength of sodium in the ratio spectrum for the in-transit data (black points). This excess absorption during transit must presumably arise in the atmosphere of \mystar\,b. Due to the similarity of target and comparison star, this simple approach allows us to verify the presence of the sodium absorption without any light curve fitting or detrending. 

To estimate the significance of the absorption signal seen in Fig.\,\ref{fig:Na_feature_flux_average}, we fitted a Gaussian to determine the in-transit flux ratio at the peak absorption and compare it to the continuum, resulting in a significance of $4.9\sigma$.

The spectra in Fig.\,\ref{fig:Na_feature_flux_average} are binned to $14$\,\AA\ wavelength bins, which is half the inherent resolution of the spectrum of $27.24 \pm 0.17$\,\AA. This value was calculated from the FWHM of target and comparison stars, as well as the instrument/grism resolution with $2\times2$ binning of 4.08\,\AA/pixel. In Fig.\,\ref{fig:Na_feature_flux_average} we see an offset of the absorption from the wavelength of the sodium doublet by one bin i.e.\ half the inherent resolution which is within the uncertainty of our wavelength solution. 

\subsection{Atmospheric Retrieval}
\label{sec:retrieval}
In order to aid interpretation 
of the transmission spectrum shown in Fig.\,\ref{fig:transmission_spectrum_both} we performed atmospheric retrievals using the \textsc{PLATON} \textsc{python} tool by \citet{Zhang2019ForwardTool, Zhang2020PLATONData}, an open-source package assuming equilibrium chemistry. 

To account for clouds and hazes \textsc{PLATON} uses three parameters: 
the logarithm of cloud-top pressure, $\log P_\textrm{cloud}$; 
the power-law index of the wavelength dependence of the scattering cross-section, $\alpha$ (with value 4 for Rayleigh scattering);
and the logarithm of a scattering slope multiplying factor, 
$\log S$.
The atmospheric opacity is thus given by $S \times \lambda^{-\alpha}$ with $\lambda$ equal to the wavelength. 

For our retrieval we considered two different models, one where the gradient of the scattering slope is fixed to Rayleigh scattering ($\alpha = 4$) and one where this parameter is free. Parameters  retrieved in all models are mass of the planet ($M_\textrm{p}$), radius of the planet at 1 bar (R$_\textrm{p, 1bar}$), the limb temperature ($T_{\textrm{limb}}$), metallicity ($\log Z/Z_{\odot}$), cloud-top pressure ($\log P_{\textrm{cloud}}$) and the scattering factor ($\log S$), which are shown along with their corresponding priors in Table~\ref{tab:retrieval_priors_and_results}. The range for the priors was mostly chosen to be wide and uniform, with the exception of the mass of the planet which was described by a Gaussian prior centred on the measured mass of the planet with standard deviation of the error (Table~\ref{tab:wasp-94_planets}). The C/O ratio was fixed to the solar value of 0.53 as we cannot constrain this value due to the lack of sensitivity to carbon and oxygen bearing species such as water or carbon monoxide in the optical wavelength range.


We chose to run the \textsc{PLATON} retrieval with a nested sampling algorithm, which is implemented in \textsc{PLATON} using the \textsc{python} package \textsc{dynesty} \citep{Speagle2020Dynesty:Evidences}. Using nested sampling allows us to compare Bayesian evidence values for the two models, describing their statistical significance. The higher the Bayesian evidence value, the better description of the data, where a difference in logarithmic evidence $\Delta \log \mathcal{Z}$ of $>5$ between two models corresponds to a 99.3~$\%$ probability \citep{Jeffreys1983TheoryProbability}, i.e.\ strongly favouring one model over the other.

\begin{figure*}
    \centering
    \includegraphics[width=\textwidth]{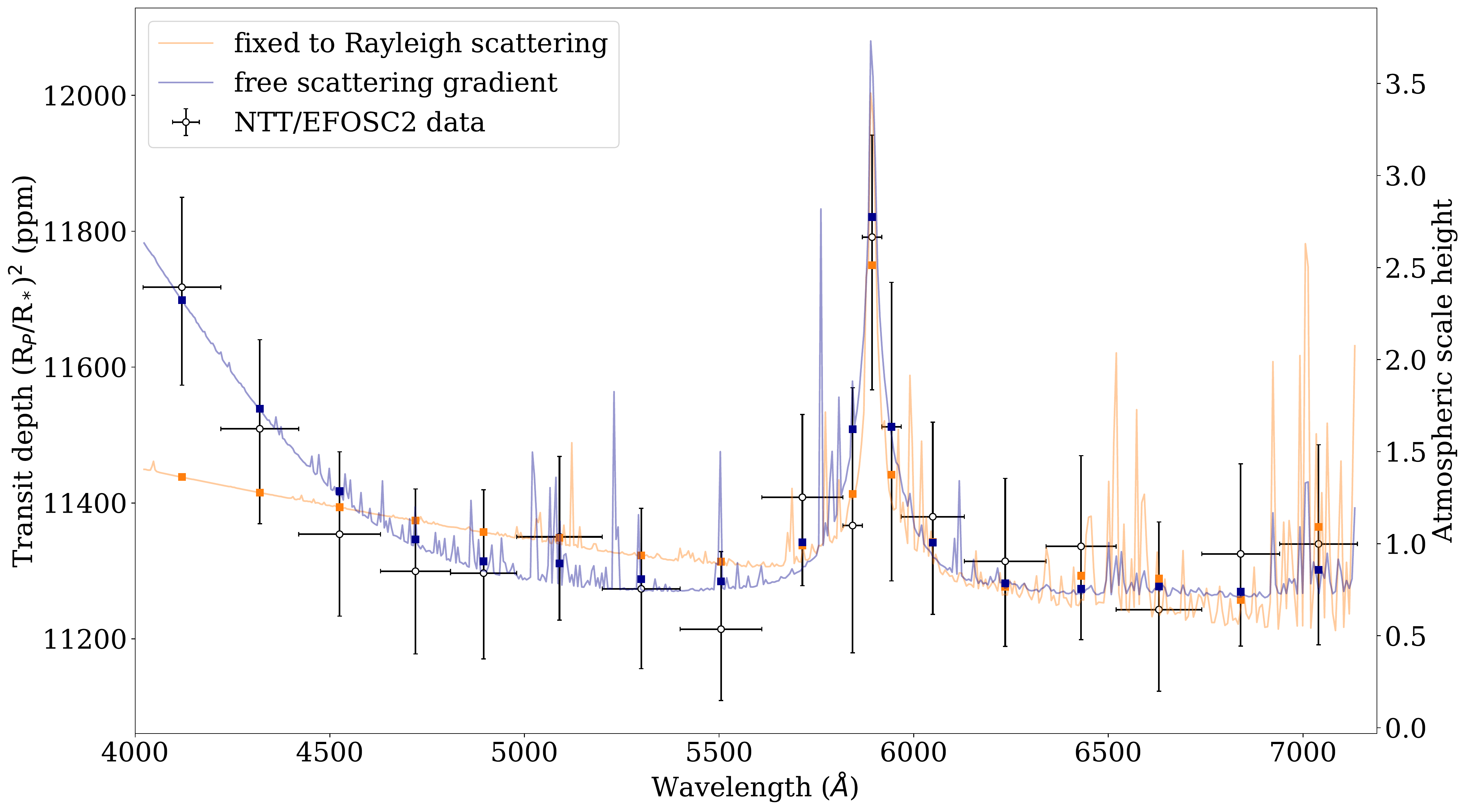}
    \caption{Obtained transmission spectrum for \mystar\ using the GP simultaneous detrending method (black) and the two \textsc{PLATON} models, one where the scattering gradient is fixed to Rayleigh scattering (orange) and one where the gradient of the scattering slope is a free parameter (blue). Note that based on the Bayesian evidence, neither of the models is significantly preferred over the other. }
    \label{fig:spectrum_with_platon}
\end{figure*}

The two resulting retrieved models are plotted in Figure~\ref{fig:spectrum_with_platon}. They are both dominated by an overall scattering slope and a prominent sodium line that exhibits significant pressure broadening.  The corresponding parameter values for these retrievals are presented in Table~\ref{tab:retrieval_priors_and_results}, which also includes the values for the Bayesian evidence. These results show that while the model including a free scattering slope resulted in a higher Bayesian evidence value, the difference of $\Delta \log \mathcal{Z} = 0.53 \pm 0.33$ does not equate to  strong enough evidence to rule out the model where the gradient is fixed to Rayleigh scattering ($\alpha = 4$). It is interesting to note that 
all fitted parameters are consistent between the two models to within $1\sigma$, except for the scattering gradient where the fitted value 
deviates from the Rayleigh slope by $1.1\sigma$.

We note that the retrieved limb temperatures of $\approx1100$~K are lower than the equilibrium temperature of $1500$~K obtained by \citet{Garhart2020Eclipses}. This is consistent with the expected terminator temperatures for such hot Jupiters \citep{Kataria2016TheSpace}. In addition, such differences between retrieved limb temperatures and equilibrium temperatures can also be explained by the single-temperature retrieval being applied to the differing temperatures and compositions of morning and evening terminators in the exoplanet's atmosphere
\citep{MacDonald2020WhyAtmospheres}.

\begin{table*}
    \centering
    \caption{Retrieved parameters for the atmospheric retrieval analysis with \textsc{PLATON} and their respective priors. Values for $T_{\textrm{eff}}$ and  $R_\textrm{p}$, $M_\textrm{p}$,  $T_\textrm{eq}$ are listed in Table~\ref{tab:wasp-94_stars} and Table~\ref{tab:wasp-94_planets}, respectively.
    The two models, one including the scattering as a free gradient and one fixed to the Rayleigh scattering ($\alpha = 4$). For both models, the C/O ratio is fixed to solar at 0.53 and the stellar radius is fixed to 1.36~$R_\odot$ (see Table~\ref{tab:wasp-94_stars}). Posterior distributions of both models in the form of a corner plot are displayed in the Appendix, Fig.\,\ref{fig:corner_platon_retrievals}.}
    \label{tab:retrieval_priors_and_results}
    \begin{tabular}{llccccc}
    \hline
    & \multicolumn{2}{c}{Priors} & \multicolumn{2}{c}{fixed to Rayleigh scattering } & \multicolumn{2}{c}{free scattering gradient } \\ \hline
    Parameter & Distribution & Limits & Median & Best Fit & Median & Best Fit \\ \hline
    Planet mass $M_\textrm{p}$ ($M_\textrm{Jup}$)& Gaussian &  $\mu = $ $M_\textrm{p}$, $\sigma=0.036$ & $0.457^{+0.030}_{-0.031}$ & 0.465 & $0.458\pm 0.030$ & 0.455 \\ 
    Planet radius $R_\textrm{p}$ ($R_\textrm{Jup}$)& Uniform &  $0.9\ \times\ $R$_\textrm{p}$, $1.1\ \times\ $R$_\textrm{p}$& $1.639^{+0.032}_{-0.050}$ &  $1.672$ & $1.645^{+0.030}_{-0.037}$ &  $1.624$\\
    Limb Temperature $T_{\textrm{limb}}$ (K)& Uniform & $0.5\ \times\ T_\textrm{eq}$, $1.5\ \times\ T_\textrm{eq}$ & $1070^{+230}_{-180}$ &  $930$  &  $1110^{+220}_{-200}$ &  $1300$\\
    Metallicity ($\log Z/Z_{\odot}$)& Uniform &  -1, 3  & $1.5^{+1.0}_{-1.6}$ &  $1.7$ &  $1.46^{+0.93}_{-1.4}$ &  $1.61$\\
    Cloud-top pressure $\log P_{\textrm{cloud}}$ (Pa)& Uniform & -0.99, 5 & $2.6^{+1.4}_{-1.7}$ &  $4.1$ &  $2.6^{+1.1}_{-1.3}$ &  $2.3$\\
    Scattering factor $\log S$& Uniform & -10, 10& $0.8^{+3.6}_{-5.5}$ &  $0.2$ &  $-2.8^{+2.9}_{-2.1}$ &  $-3.9$\\
    Scattering gradient $\alpha$& Uniform & -4, 20 & \multicolumn{2}{c}{fixed to 4 } &  $13.5^{+4.4}_{-8.7}$ &  $18.8$\\
    \hline
    Bayesian evidence $\log \mathcal{Z}$ && &  \multicolumn{2}{c}{$131.40 \pm 0.26$} & \multicolumn{2}{c}{$131.93 \pm 0.29$} \\
    \hline
    \end{tabular}
    
\end{table*}

\subsection{The width of the sodium line}
\label{sec:line_width}
\begin{figure}
    \centering
    \includegraphics[width=\columnwidth]{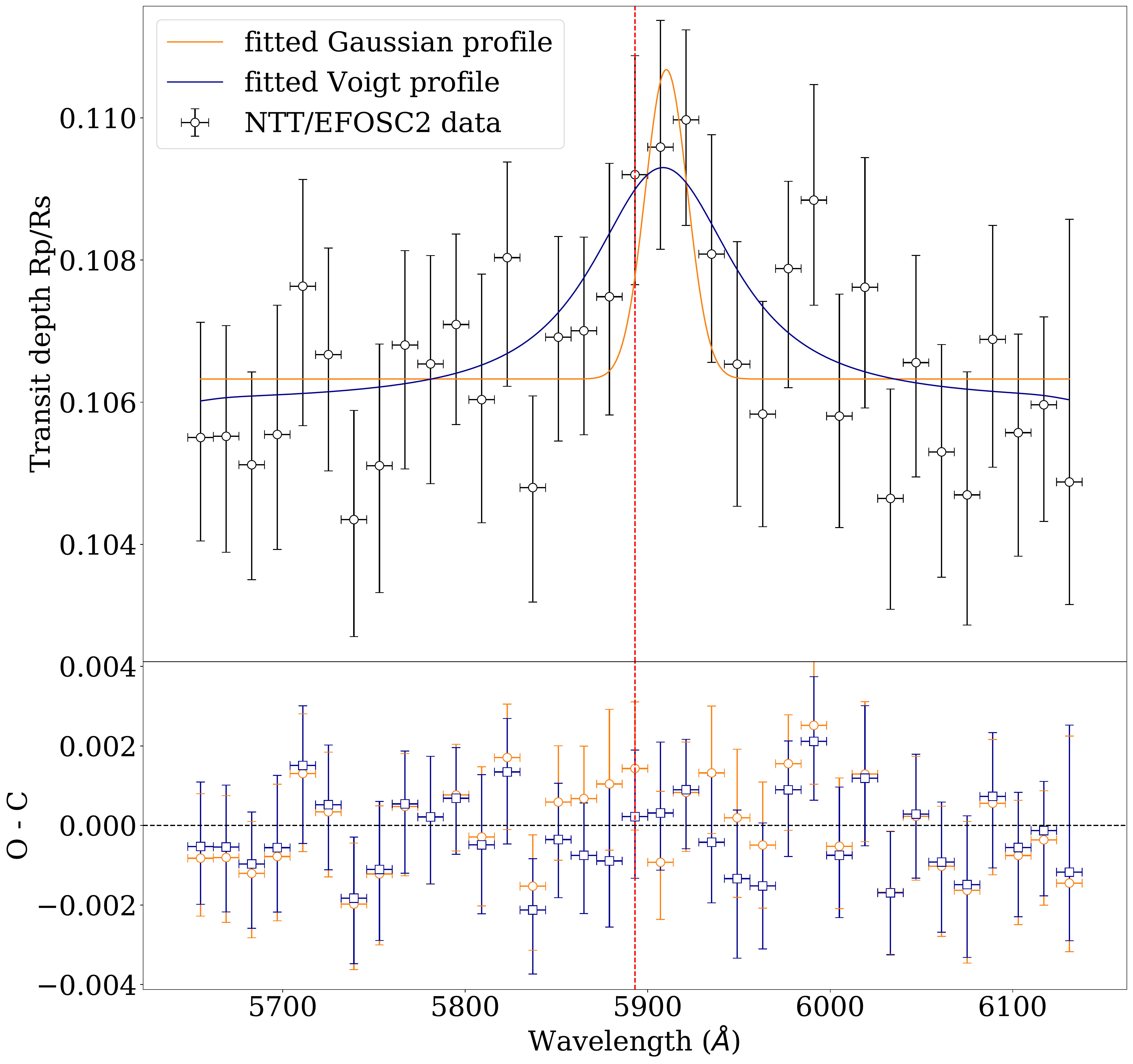}
    \caption{Transmission spectrum of \mystar\,b when sampling at Nyquist rate (14~\AA\ bin width) between $5600-6200$~\AA\ with two competing models and corresponding residuals. The blue line corresponds to the model solely considering instrumental inherent broadening, while the orange one fits for an extra width equal to a Voigt profile (convolution of the instrumental profile with a Lorentzian profile i.e.\ the pressure-broadened sodium absorption). The expected central wavelength  of the NaI doublet is indicated with the red dashed line. }
    \label{fig:Na_feature_models}
\end{figure}

Our atmospheric retrieval shows significant broadening in the sodium line, corresponding to pressure broadening in a relatively cloud-free atmosphere (Fig.\,\ref{fig:spectrum_with_platon}). The two wide data points neighbouring the three narrow bins around the sodium line are also elevated, suggesting the pressure broadening may be significantly detected in our observed transmission spectrum. 
To investigate the sodium line profile further, we split the spectrum
between $5600 - 6200$~\AA\ into 35 wavelength bins, each with a width of $14$~\AA\ (the Nyquist sampling, see Section~\ref{sec:Nyquist}). The middle bin was centred on the sodium doublet. The spectroscopic light curves were then fitted using linear detrending, resulting in a higher-resolution transmission spectrum that is plotted in Fig.~\ref{fig:Na_feature_models}. 

The sodium line remains clearly visible in this higher-resolution transmission spectrum, and the line profile was fitted with two different models: a Gaussian with FWHM fixed to the instrumental resolution (27.24\,\AA) and a Voigt profile with an unconstrained additional width. The latter model is a convolution of the former, instrumental model, with a Lorentzian describing the pressure-broadened sodium described by the fitting parameter $\gamma$. The FWHM of the Lorentzian profile is $2\gamma$. To account for uncertainties in our wavelength solution we allowed both models to fit for the centre of the sodium absorption. The two model fits are included in Fig.~\ref{fig:Na_feature_models}. 


The instrumental width does not 
fit the sodium line well and suggests broadening is needed. We find a best fitting FWHM for the Voigt profile of $78_{-32}^{+67}$\,\AA, where the FWHM of the Lorentzian component is $68_{-38}^{+72}$\,\AA, which is inconsistent with zero at the $1.8\sigma$ level. 
The Bayesian evidence for the Voigt model, however, is lower than the value for the instrument-only broadening model by $\Delta \mathcal{Z} = 1.69 \pm 0.15$.  

\section{Discussion}
\label{sec:discussion}

\subsection{Precision of observations}
In this paper we present the first transmission spectroscopy measurements with ESO's EFOSC2 instrument on the 3.6\,m NTT. EFOSC2 is a grism spectrograph, with a simple optical design, which is favourable for high precision observations. Our wavelength-dependent light curves shown in Figs.\,\ref{fig:nonGP_all_lightcurves}\,\&\,\ref{fig:all_fitted_lightcurves_GP} are relatively free of systematic noise, with the RMS amplitude of residuals only $1.1\times$ and $1.4\times$ above the expected photon noise for each of the detrending approaches (Sections\,\ref{sec:poly}\,\&\,\ref{sec:gp}). 
This is better than typically seen with larger 8--10\,m class telescopes, where the improved photon noise often cannot be exploited because the precision is limited by systematics. Consequently, transmission spectroscopy with medium-sized telescopes such as the NTT  can achieve comparable precision. 

In our transmission spectra of Figs.\,\ref{fig:transmission_spectrum_both}\,\&\,\ref{fig:spectrum_with_platon} we achieve an average absolute precision of 128\,ppm in 200\AA\ bins, which is less than half the atmospheric scale height of our target planet \mystar\,b. This precision compares very favourably with single-transit transmission spectroscopy in low-resolution carried out with much larger telescopes such as the 8.2\,m VLT using the FORS2 instrument \citep[e.g.\ 200 ppm for 150\AA\ wide bins in][]{Wilson2020Ground-basedWASP-103b}, as well as from space with HST using the STIS instrument \citep[e.g.\ 221 ppm for $\sim$ 175\AA\ wide bins with grism G750L or 175 ppm for $\sim$ 55\AA\ wide bins with grism G430L in][]{Sheppard2021TheHAT-P-41b}. 

In part, the very high precision of our single-transit observations arises from the binary nature of the system, with WASP-94B providing an ideal comparison star of very similar brightness and spectral type at an ideal separation of just 15\,arcsec from the target \citep[similar to the situation for XO-2\,b, e.g.][]{Pearson2019Ground-basedCalibration}. We note that this makes \mystar\,b particularly well suited to a future intensive ground-based programme of atmospheric characterisation. 

\begin{figure}
    \centering
    \includegraphics[width = \columnwidth]{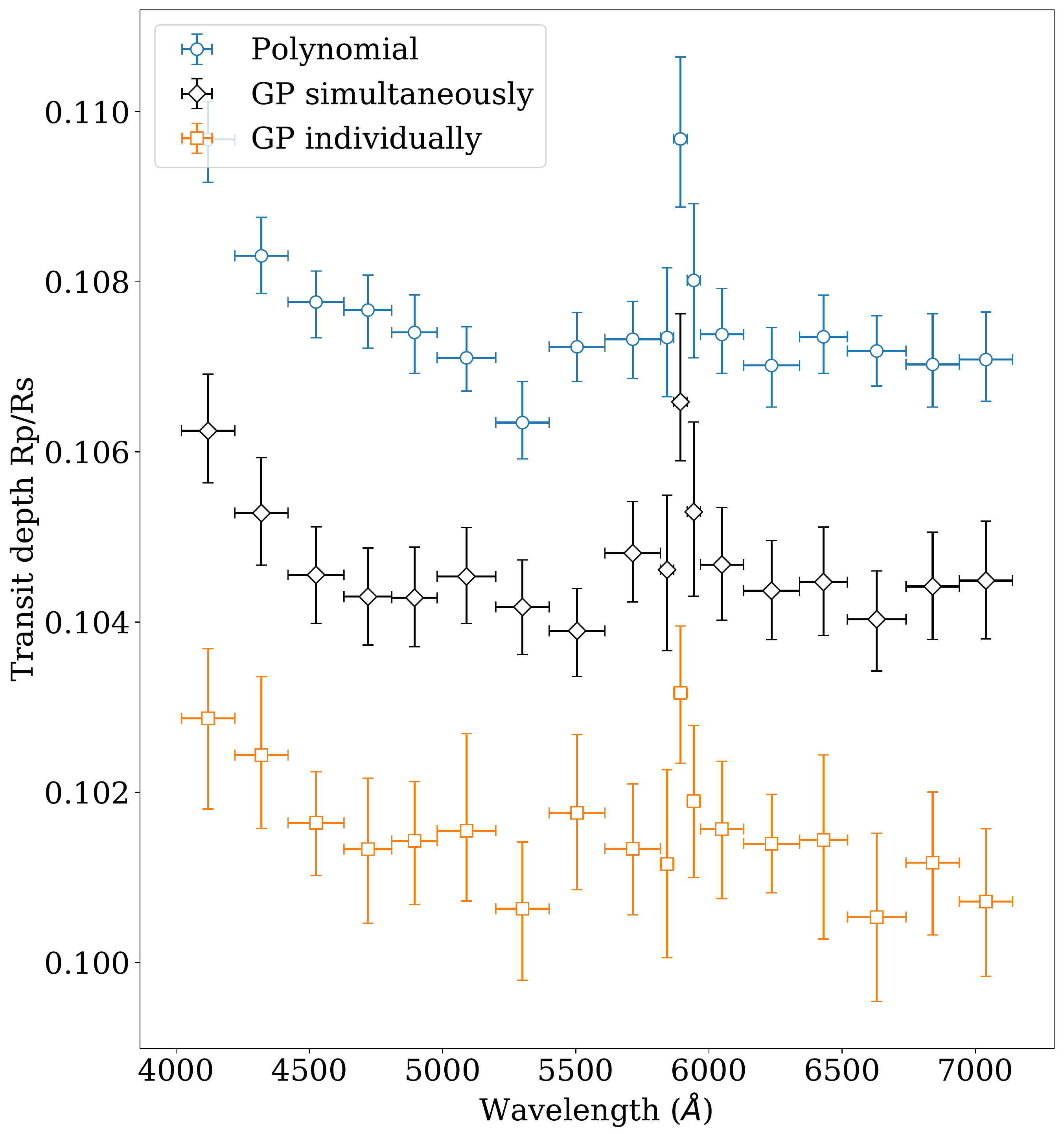}
    \caption{\mystar\,b transmission spectra comparing the precision achieved with three different detrending methods. From top to bottom:
    first order polynomial detrending is used for the individual light curves ($98 \pm 3$ ppm precision in $200$\,\AA\ wide bins); GP detrending simultaneously on all spectroscopic light curves with shared GP hyperparameters for length scales ($128 \pm 2$ ppm precision); GP detrending with all bins fitted independently ($181 \pm 6$ ppm precision).}
    \label{fig:transmission_spectra_different_detrending}
\end{figure}

We also found the absolute precision in the transmission spectrum with the GP detrending was improved by fitting all light curves simultaneously while linking the GP hyper-parameters for length-scale.
This allowed us to account for systematic features in the light curves while imposing similar detrending. The precision is significantly lower when fitting the light curves with GPs individually, with an average precision of 181\,ppm compared to 128\,ppm for bins $\sim 200$\,\AA\ wide. We illustrate this in Fig.~\ref{fig:transmission_spectra_different_detrending}, where we show the different transmission spectra with the three different detrending methods, from top to bottom with increasing uncertainties. 

\subsection{Sodium absorption}
Our analysis reveals sodium to be present in the atmosphere of \mystar\,b at a significance of $4.9\sigma$. The absorption feature is visible in transmission spectra measured with each of our detrending models (linear polynomial and GP; Fig.\,\ref{fig:transmission_spectrum_both}). We also confirmed the detection and significance of sodium absorption independently of our light curve fitting using a direct comparison of average spectra in and out of transit (Fig.\,\ref{fig:Na_feature_flux_average}). This was possible because of the very close match between the stellar spectra of the target star \mystar\ and its binary companion, WASP-94B (Fig.\,\ref{fig:wavelength_bins})

Our measured transmission spectrum and its retrieval models using \textsc{PLATON} in Fig.\,\ref{fig:spectrum_with_platon} suggest a pressure-broadened sodium absorption line arising in a relatively cloud-free atmosphere. To investigate 
the sodium line profile further, we extracted a second
transmission spectrum using the full spectral resolution available from our data (extracted with the Nyquist sampling of 14\,\AA\ bins; Fig.\,\ref{fig:Na_feature_models}). Our best-fitting line profile is indeed broader than the instrumental resolution, with a FWHM of $78^{+67}_{-32}$\AA\ (see Section\,\ref{sec:line_width} and Fig.\,\ref{fig:Na_feature_models}), although at this spectral resolution the signal-to-noise ratio of our single-transit observation is low and Bayesian model comparison does not support the additional free parameter of the broadened model. Further observations are required to determine whether sodium absorption in the transmission spectrum of \mystar\,b is indeed significantly pressure-broadened. 


Detections of sodium absorption in hot Jupiters remain relatively rare, especially in the low-resolution regime, with less than ten cases from HST and only a handful measured from the ground \citep[e.g.][]{Sing2012GTCSpectroscopy,Nikolov2016VLTGROUND,Nikolov2018AnExoplanet,Chen2020DetectionAtmosphere}. In some cases the detections are disputed, for instance in the original case of HD\,209458\,b \citep{Casasayas-Barris2020IsStudies, Casasayas-Barris2021TheESPRESSO}. In most cases the detected sodium lines are consistent with being narrow, and arising from relatively high in the planetary atmosphere, with the pressure-broadened absorption from the lower atmosphere being masked by condensates \citep[e.g.][]{Nikolov2014HSTHAT-P-1b,Sing2016ADepletion}. Only in a few exoplanet atmospheres the pressure-broadening has been observed so far, such as  WASP-39\,b \citep{Fischer2016WASP-39b},WASP-96\,b \citep{Nikolov2018AnExoplanet} and WASP-62\,b \citep{Alam2021EvidenceZone}. If confirmed with higher signal-to-noise observations, our tentative detection of pressure-broadened sodium absorption in \mystar\,b would indicate a relatively cloud-free atmosphere, making it a priority target for infra-red observations aiming to detect molecular bands and measure atmospheric abundances.



\subsection{Scattering slope}
\label{sec:slope}
\begin{figure}
    \centering
    \includegraphics[width=\columnwidth]{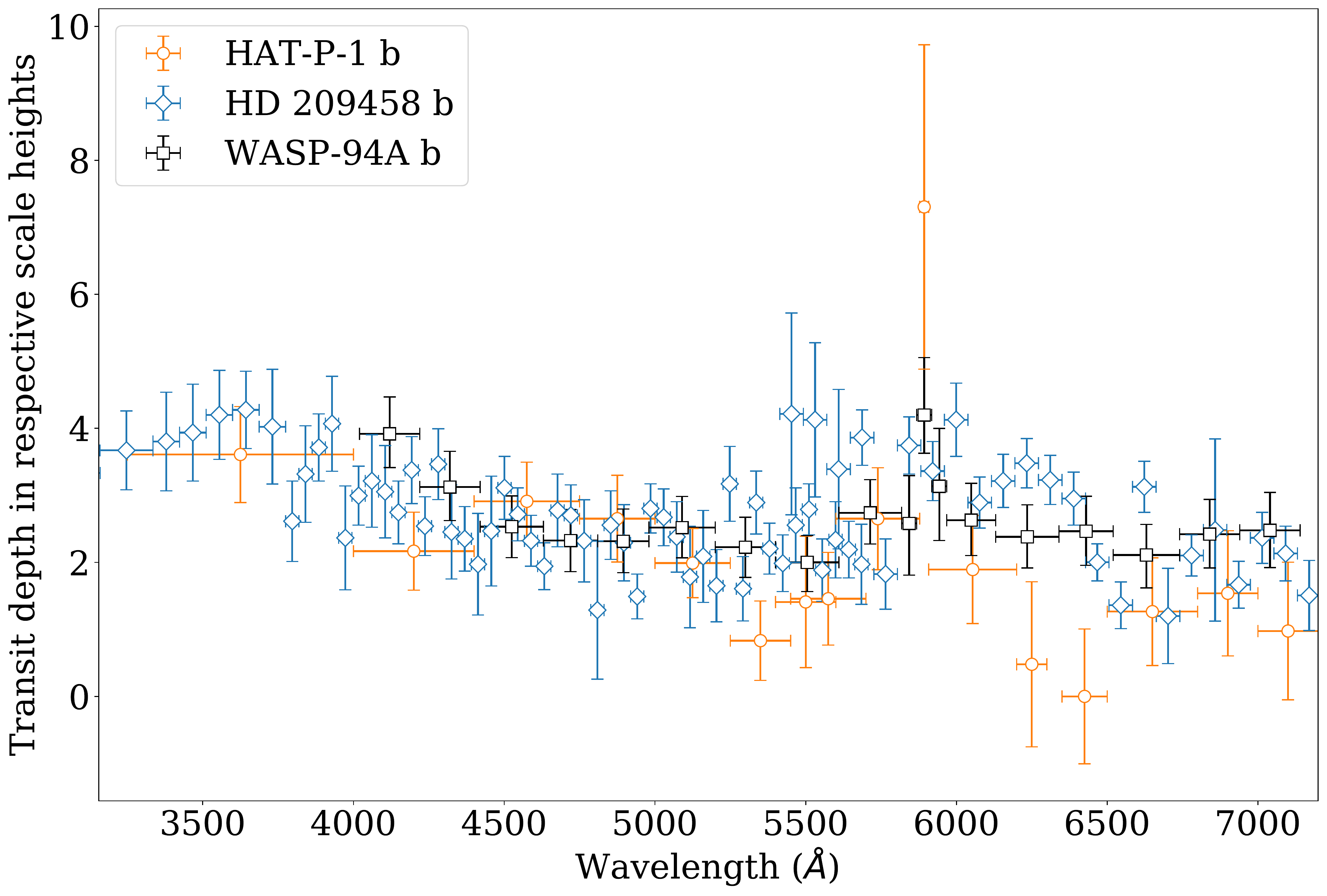}
    \caption{Transmission spectrum of \mystar\,b (black) in comparison to the transmission spectra of HD\,209458\,b (blue) \citep{Sing2016ADepletion} and HAT-P-1\,b (orange) \citep{Nikolov2014HSTHAT-P-1b} as a function of scale height. One atmospheric scale height corresponds to $262$\,ppm for \mystar\,b, $166$\,ppm for HD\,209458\,b and $146$\,ppm for HAT-P-1\,b. Both HD\,209458\,b and HAT-P-1\,b have similar properties to \mystar\,b and orbit stars with similar spectral type (see text for details). }
    \label{fig:spectra_comparison_P1_HD209}
\end{figure}

Our transmission spectra show a clear and steep upward slope at the blue end of the spectrum (Figs.\,\ref{fig:transmission_spectrum_both}\,\&\,\ref{fig:spectrum_with_platon}), which is indicative of Rayleigh scattering in the planetary atmosphere. Similar slopes have been observed for alike planets such as HD\,209458\,b \citep[][]{Sing2016ADepletion} and HAT-P-1\,b \citep[][]{Nikolov2014HSTHAT-P-1b}, demonstrated in Fig.~\ref{fig:spectra_comparison_P1_HD209}. Both exoplanets possess close resemblance to \mystar\,b in physical properties, while also orbiting similar stars (G0 type stars). HD\,209458\,b has a 3.52-day period, with a mass of 0.69~$M_{\textrm{Jup}}$, a radius of 1.38~$R_{\textrm{Jup}}$ \citep{Bonomo2017ThePlanets} and equilibrium temperature of $1480$~K \citep[e.g.\ ][]{Evans2015AModels}, while HAT-P-1\,b orbits its host star every 4.47~days, with a mass of 0.53~$M_{\textrm{Jup}}$, a radius of 1.32~$R_{\textrm{Jup}}$ and equilibrium temperature of $1320$~K \citep{Nikolov2014HSTHAT-P-1b}.

We find 
evidence 
for a super-Rayleigh slope in the transmission spectrum (see Section\,\ref{sec:retrieval}). The strength of this evidence depends of the choice of detrending method, which in turn sets the precision of the transmission spectrum (Fig.\,\ref{fig:transmission_spectra_different_detrending}).
With the smaller uncertainties estimated with linear detrending, we find reasonably strong evidence of a super-Rayleigh slope ($\Delta \log \mathcal{Z} = 3.70 \pm 0.71$ which corresponds to $2-2.5\sigma$ significance). However, using the more conservative uncertainties from the Gaussian Process detrending we find the evidence for a super-Rayleigh slope is 
weaker 
($\Delta \log \mathcal{Z} = 0.54 \pm 0.38$ which is $\lesssim 1 \sigma$).

Super-Rayleigh slopes have been 
observed in several exoplanet atmospheres before \citep[e.g.\ HD189733\,b, WASP-127\,b, WASP-21\,b, WASP-74\,b, WASP-104\,b;][]{Sing2011GranSpectrophotometry, Palle2017Feature-richWASP-127b, Alderson2020LRG-BEASTS:WASP-21b, Luque2020ObliquitySystem, Chen2021AnWASP-104b}. Several explanations have been suggested for this, including based on additional opacity sources 
such as mineral clouds \citep[e.g.\ ][]{LecavelierDesEtangs2008Rayleigh189733b}, photochemical hazes --- especially for exoplanets like \mystar\,b with equilibrium temperatures around $1000 - 1500$~K \citep[][]{Kawashima2019TheoreticalTemperature,Ohno2020Super-RayleighHaze}, a combination of clouds and haze layers \citep[e.g. ][]{Pont2013TheObservations} or alternatively stellar activity \citep[e.g. ][]{McCullough2014WaterTransit, Espinoza2019ACCESS:Magellan/IMACS, Weaver2020ACCESS:K}. Thermal inversion could potentially push a H$_2$ Rayleigh scattering opacity source to a steeper slope \citep{Fortney2008AAtmospheres}, however, planets below $\sim 2000$\,K are less likely to have thermal inversions due to species such as TiO and VO as they are not expected to be in the gas phase at these temperatures \citep[e.g.][]{Gandhi2019NewJupiters} and given the \mystar\,b's equilibrium temperature of $1500$\,K, a thermal inversion on the dayside atmosphere is unlikely.

In the case of \mystar\ we do not expect stellar activity to play a significant role as the effects of spots and faculae were estimated to be $<30$~ppm for F8-type stars assuming $1\%$ transit depth \citep{Rackham2019TheStars}, which is not detectable with our precision and cannot account for the detected slope. In addition, in the discovery paper by \citet{Neveu-Vanmalle2014WASP-94System} there was no mention of any photometric variations in the lightcurve of \mystar\,b. Nevertheless, we tested this case by running retrieval models with \textsc{PLATON} including stellar activity instead of the gradient of the scattering slope. These retrievals included fitting for spot temperature and spot coverage fraction, which was limited to $<$7$\%$ as estimated for F8-type stars by \citet{Rackham2019TheStars}. These retrievals confirmed that the stellar activity was not able to account for the observed blueward slope. 

\subsection{Evidence for atmospheric escape}

As part of our investigation into stellar activity as a potential explanation of the steep scattering slope (Section\,\ref{sec:slope})
we checked
whether there is chromospheric emission in the CaII H\&K lines using high-resolution observations with the HARPS spectrograph \citep{Mayor2003SettingHARPS} of \mystar\footnote{Based on observations collected at the European Southern Observatory under ESO programme 097.C-1025(B).}.   

\begin{figure}
    \centering
    \includegraphics[width=\columnwidth]{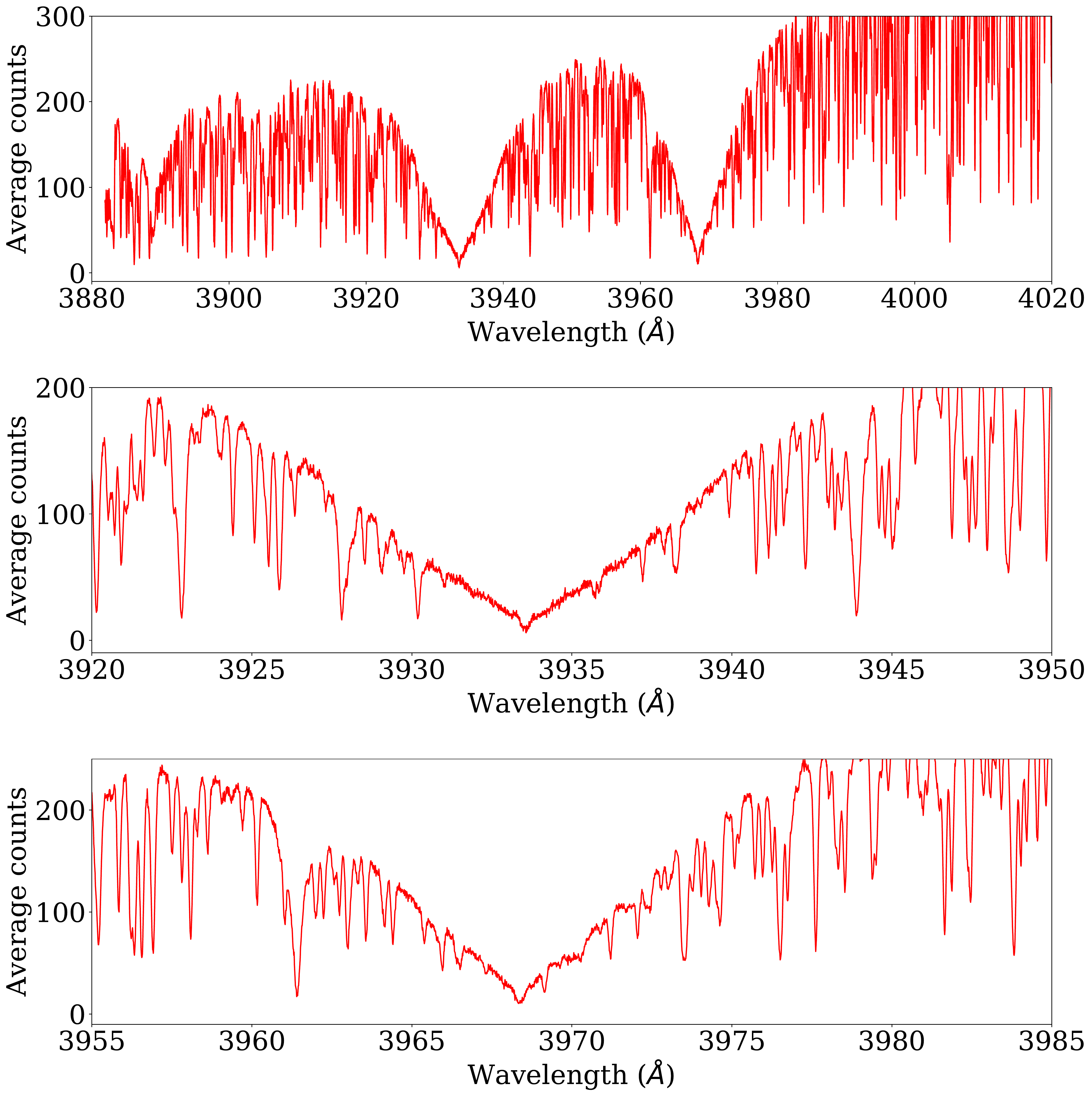}
    \caption{High-resolution HARPS spectrum of WASP-94A: Both CaII H\&K lines (top) and zoomed in on the two lines (middle: K line with centre at 3933.66\,\AA, bottom: H line with centre at 3968.47\,\AA). There is no emission seen in either of the core lines, instead there is a dip visible, resulting in a \lrhk value of $-5.18  \pm 0.27$.}
    \label{fig:CaHK_line}
\end{figure}

Following \citet{Lovis2011TheVelocities} and using calibrations by \citet{Noyes1984RotationStars} and \citet{Middelkoop1982MagneticStars} we calculated the \lrhk\ of \mystar\ to be $\lrhk = -5.18 \pm 0.27$. This value is below $-5.1$ which corresponds to a basal level exhibited by inactive FGK stars and we found that the H\&K lines do not show emission in the core.  Instead, the H\&K lines appear to have excess narrow absorption in their core, which can be seen in  Fig.~\ref{fig:CaHK_line}. This excess absorption is similar to that seen in WASP-12\,b and a number of other hot Jupiters \citep{Fossati2013AbsorbingSystem}, where it is interpreted as absorption by circumstellar material escaping the hot Jupiter atmosphere and absorbing the stellar emission of CaII in the H\&K lines. This interpretation is  supported by the statistical analysis of a larger sample of stars by \citet{Haswell2020DispersedStars}. It is also supported by the fact that a \lrhk\ value this low (and in combination with the colour of the star $B-V = 0.69$) would suggest a rotation period of $ 40$~days and a high age of $8$~Gyr \citep{Mamajek2008ImprovedDiagnostics}.
The rotation period was determined by \citet{Neveu-Vanmalle2014WASP-94System} to be $19.5$\,days, based on their measurement of $v \sin i_*$, the projected stellar rotational velocity. We conclude that \mystar\ is most likely enshrouded in gas escaping from its hot Jupiter. 


\section{Conclusions}
\label{sec:conclusions}
We have presented a  low-resolution optical transmission spectrum of the hot Jupiter \mystar\,b: the first atmospheric characterisation of this highly-inflated exoplanet. Our observations form part of the LRG-BEASTS survey and made use of the EFOSC2 instrument at the ESO NTT, which has not been utilised for transmission spectroscopy previously. Our transmission spectrum from one full transit of \mystar\ covers the wavelength range $4020-7140$~\AA, with an average transit-depth precision of $128$\,ppm for $\sim 200$\,\AA\ wide bins. This corresponds to less than half a scale height in the planetary atmosphere. The binary companion WASP-94B served as an ideal comparison star with very similar 
brightness and spectral type to \mystar,
resulting in an average RMS-noise to expected photon noise ratio of $1.15$.

We compared two methods for accounting for systematic trends in our light curves, 
in order to show that our results are insensitive to the choice of detrending model. In the first case we utilized a simple linear detrending approach, while in the second we modelled the noise using a Gaussian Process. For the Gaussian Process, we employed a kernel using sky background and telescope derotator angle as kernel inputs. We also simultaneously fitted all spectroscopic light curves, linking the Gaussian Process hyperparameters for length scales. While this was computationally expensive, we found this approach provided more consistent detrending between wavelength bins and an overall improvement in the precision of transmission spectrum (from an average of 181\,ppm to 128\,ppm). 

Our subsequent analysis of the transmission spectrum reveals sodium absorption in the atmosphere of \mystar\,b with a significance of 4.9$\sigma$. The sodium absorption is visible even 
in a direct ratio of in- and out-of-transit spectra, and we find evidence for substantial pressure broadening.
Our detection of strong sodium absorption with probable pressure broadening
indicates 
a clear atmosphere.

In addition, our transmission spectrum shows a steep blueward slope at short wavelengths. 
Although our retrievals are consistent with Rayleigh scattering, the best-fitting slope is much steeper. 
A super-Rayleigh slope might be caused by mineral clouds or photochemical hazes.

Finally, we note that the relatively clear atmosphere of \mystar\,b combined with availability of an ideal comparison star makes it a prime target for further atmospheric characterisation, and especially abundance measurements in the infrared.

\section*{Acknowledgements}
The authors are thankful to the anonymous reviewer for valuable feedback that helped improve this manuscript.
Based on observations collected at the European Southern Observatory, under ESO programmes 099.C-0390(A) and 097.C-1025(B). PJW acknowledges support from the UK Science and Technology Facilities Council (STFC) under consolidated grants ST/P000495/1 and ST/T000406/1, and SG acknowledges STFC support under grant  ST/S000631/1. This research has made use of the NASA Exoplanet Archive, which is operated by the California Institute of Technology, under contract with the National Aeronautics and Space Administration under the Exoplanet Exploration Program. We acknowledgement the use of the ExoAtmospheres database during the preparation of this work. 

\section*{Data Availability}
The raw data used in our analysis are available from the ESO data archive. The reduced light curves presented in this article will be available via VizieR at CDS.



\bibliographystyle{mnras}
\bibliography{references.bib} 




\appendix

\section{Posterior plots}

\begin{figure*}
     \centering
     \includegraphics[width=\textwidth]{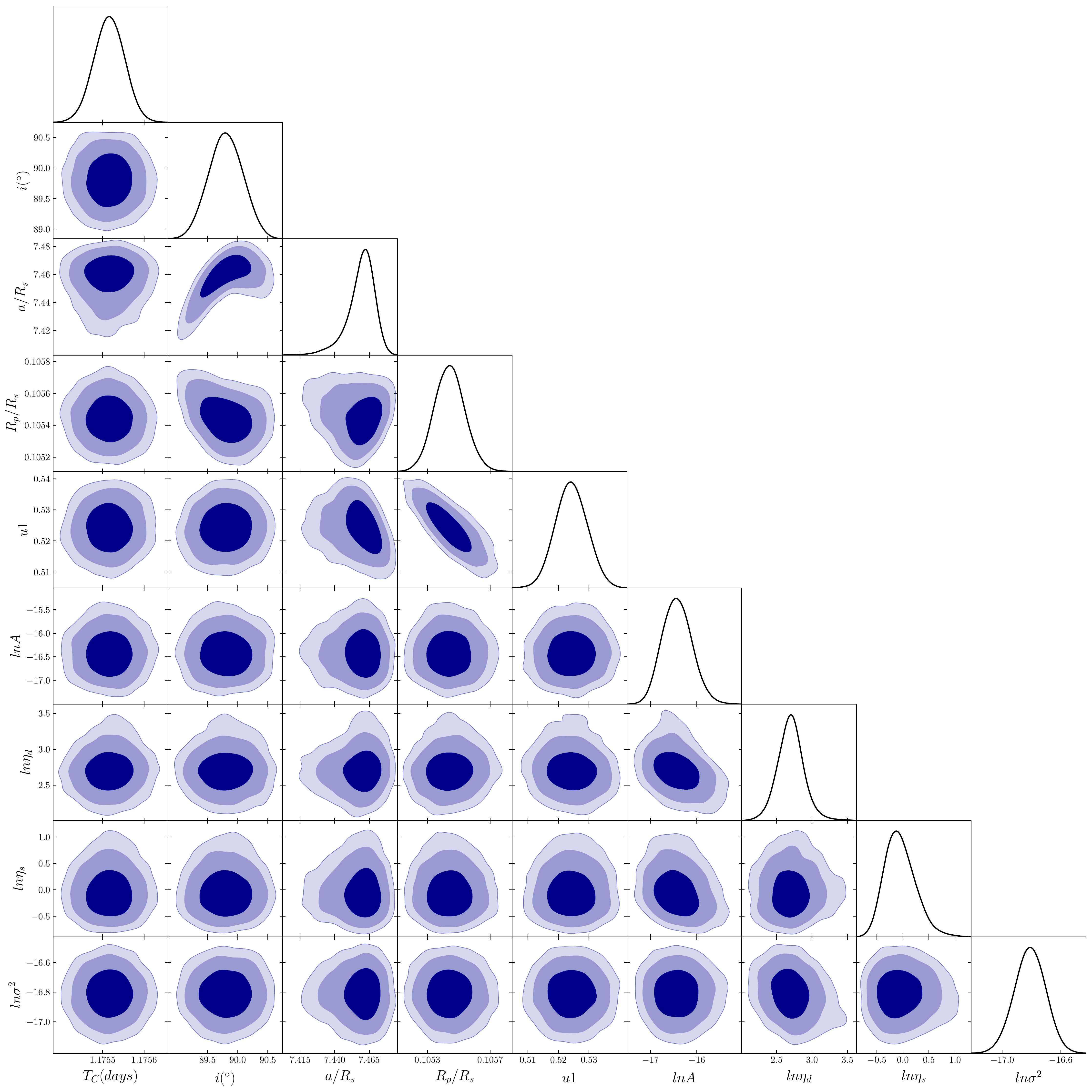}
     \caption{Corner plot of the posterior distributions of parameter values as determined by the white-light light curve fit using a GP to model systematics. Some of the parameters describe the transit: time of mid-transit $T_C$, inclination $i$, scaled stellar radius $a/R_s$, transit depth $R_p/R_s$ and limb-darkening coefficient $u1$; while the GP is described by the amplitude $A$, the logarithmic inverse length scales $\ln \eta_d$ and $\ln \eta_s$ and the variance of the white noise kernel $\ln \sigma^2$. All retrieved parameters and their prior distributions are summarised in Table\,\ref{tab:system_parameters} and further described in the main text of the paper. }
     \label{fig:corner_GP_WL}
 \end{figure*}

\begin{figure*}
     \centering
     \includegraphics[width=\textwidth]{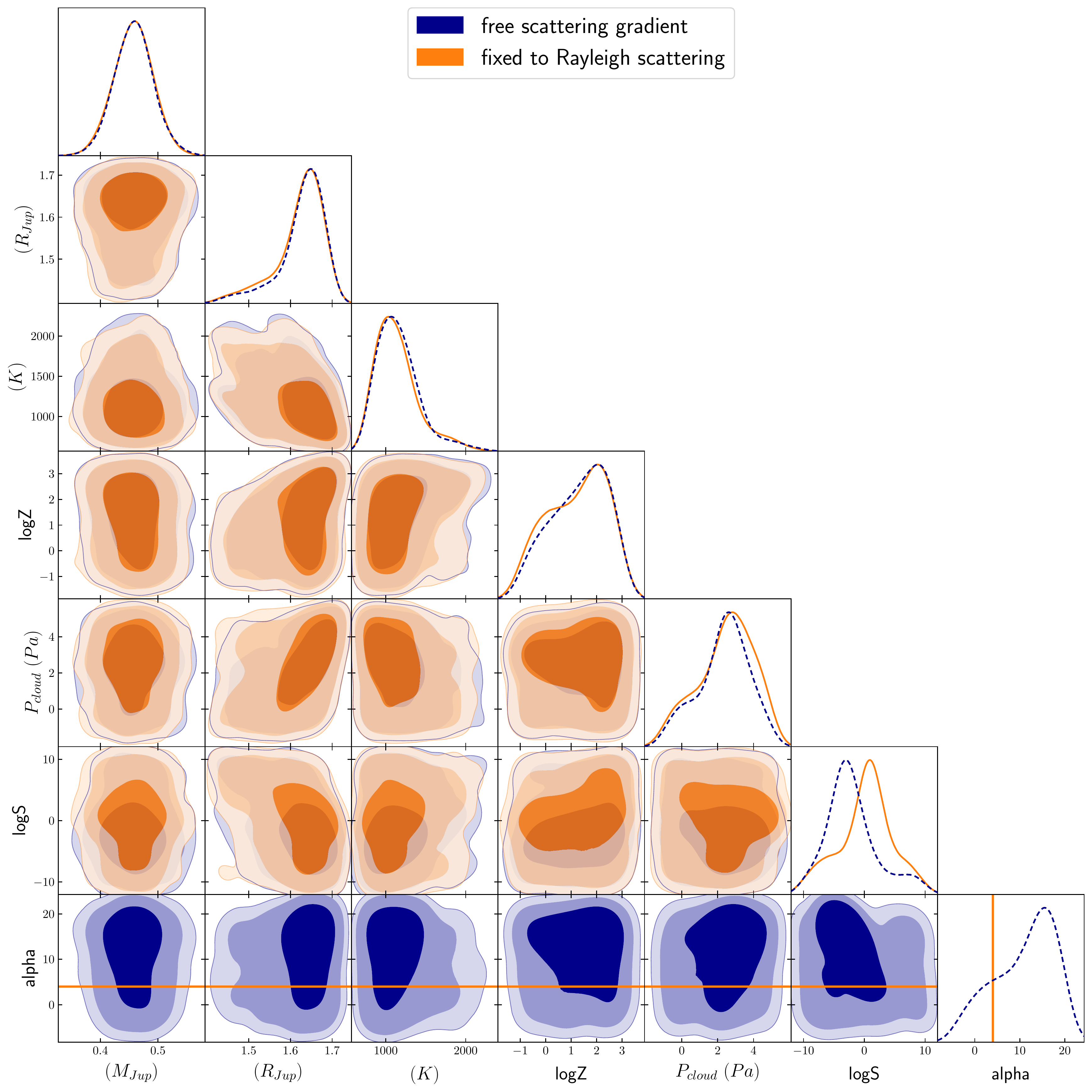}
     \caption{Posterior distributions of the parameters fitted by the two atmospheric retrieval models using \textsc{PLATON} as shown in Fig.\,\ref{fig:spectrum_with_platon} with all retrieved values and uncertainties displayed in Table\,\ref{tab:retrieval_priors_and_results}. Following the same colour code, blue represents the retrieved parameters fitted by the free scattering gradient model and orange describes the distributions of the parameters when the gradient was fixed at Rayleigh scattering i.e.\ $\alpha = 4$.}
     \label{fig:corner_platon_retrievals}
 \end{figure*}


\bsp	
\label{lastpage}
\end{document}